\documentclass[12pt,doublecolumn]{IEEEtran}
\usepackage{makeidx}
\usepackage{amsfonts}
\usepackage{amsmath}
\usepackage{cite}
\usepackage{url}
\usepackage{epsfig,dsfont,amssymb}
\usepackage{wrapfig,floatflt}
\usepackage{theorem}
\usepackage{graphics}
\usepackage{subfig}
\usepackage{stfloats}
\usepackage{mathrsfs}
\usepackage{color}
\usepackage{mathtools}
\usepackage{hyperref}
\definecolor{gray}{RGB}{128,128,128}

\def \bs {\boldsymbol}
\def \Pr {\mathbb{P}}

\def \tr {\mathrm{trace}}

\def \rank {\mathrm{rank}}
\def \vec {\mathrm{vec}}

\def \A {\mathcal{A}}

\def \I {\mathrm{I}}
\def \R {\mathbb{R}}

\def \sF {\mathcal{F}}
\def \rmF {\mathrm{F}}

\def \y {\bs y}
\def \x {\bs x}

\def \hX {\hat{X}}
\def \a {\bs a}
\def \z {\bs z}
\def \w {\bs w}

\def \g {\bs g}
\def \H {\mathcal{H}}

\def \r {\bs r}
\def \u {\bs u}
\def \v {\bs v}

\def \N {\mathcal N}
\def \E {\mathbb E}

\def \df {\stackrel{\mathrm{def}}{=}}

\def \nn {\nonumber}

\def \mathbbm {\mathbb}
\def \tmmathbf {\boldsymbol}

\newcommand{\tmop}{\operatorname*}

\newtheorem{lm}{Lemma}
\newtheorem{thm}{Theorem}
\newtheorem{cor}{Corollary}
\newtheorem{pro}{Proposition}
\newtheorem{defi}{Definition}
\newtheorem{remark}{Remark}

\title{The Stability of Low-Rank Matrix Reconstruction:\\ a Constrained Singular Value View}

\author{Gongguo~Tang,~\IEEEmembership{Member,~IEEE,} and Arye~Nehorai,~\IEEEmembership{Fellow,~IEEE}\thanks{The authors are with the Preston M. Green Department of Electrical \& Systems Engineering, Washington University in St. Louis, St. Louis, MO, USA (Emails: \{gt2, nehorai\}@ese.wustl.edu).}\thanks{This work was supported by ONR Grant N000140810849, and NSF Grants CCF-1014908
and CCF-0963742.}\\
}
\begin{document}

\maketitle
\noindent
\begin{abstract}
The stability of low-rank matrix reconstruction with respect to noise is investigated in this paper. The $\ell_*$-constrained minimal singular value ($\ell_*$-CMSV) of the measurement operator is shown to determine the recovery performance of nuclear norm minimization based algorithms. Compared with the stability results using the matrix restricted isometry constant, the performance bounds established using $\ell_*$-CMSV are more concise, and their derivations are less complex. Isotropic and subgaussian measurement operators are shown to have $\ell_*$-CMSVs bounded away from zero with high probability, as long as the number of measurements is relatively large. The $\ell_*$-CMSV for correlated Gaussian operators are also analyzed and used to illustrate the advantage of $\ell_*$-CMSV compared with the matrix restricted isometry constant. We also provide a fixed point characterization of $\ell_*$-CMSV that is potentially useful for its computation.
\end{abstract}

\vspace{-0.3cm}
\section{Introduction}
\label{sec:intro}
\noindent
The last decade witnessed the burgeoning of exploiting low dimensional structures in signal processing, most notably the sparseness for vectors \cite{Candes2006NearOptimal, Donoho2006Compressed}, low-rankness for matrices \cite{Recht2010Nuclear, fazel2002matrix, candes2009lowrank}, and low-dimensional manifold structure for general non-linear data sets \cite{Tenenbaum2000Isomap, roweis2000nonlinear}. This paper focuses on the stability problem of low-rank matrix reconstruction. Suppose $X \in \R^{n_1\times n_2}$ is a matrix of rank $r \ll \min\{n_1,n_2\}$, the low-rank matrix reconstruction problem aims at recovering matrix $X$ from a set of linear measurements $\y$ corrupted by noise $\w$:
\begin{eqnarray}\label{intro:model}
  \y = \A(X) + \w,
\end{eqnarray}
where $\A: \R^{n_1\times n_2} \rightarrow \R^m$ is a linear measurement operator. Since the matrix $X$ lies in a low-dimensional sub-manifold of $\R^{n_1\times n_2}$, we expect $m \ll n_1n_2$ measurements would suffice to reconstruct $X$ from $\y$ by exploiting the signal structure. Application areas of model \eqref{intro:model} include factor analysis, linear system realization \cite{ghaoui1993rank, fazel2001a}, matrix completion \cite{recht2009exact, Candes2009Completion}, quantum state tomography \cite{gross2009quantum}, face recognition \cite{Basri2003lambertian, Candes2011PCA}, Euclidean embedding \cite{linial1995the}, to name a few (See \cite{Recht2010Nuclear, fazel2002matrix, candes2009lowrank} for discussions and references therein).

Several considerations motivate the study of the stability of low-rank matrix reconstruction. First, in practical problems the linear measurement operator $\A$ is usually used repeatedly to collect measurement vectors $\y$ for different matrices $X$. Therefore, before taking the measurements, it is desirable to know the goodness of the measurement operator $\A$ as far as reconstructing $X$ is concerned. Second, a stability analysis would offer means to quantify the confidence on the reconstructed matrix $X$, especially when there is no other ways to justify the correctness of the reconstructed signal. 

In the current work, we define the $\ell_*$-constrained minimal singular value ($\ell_*$-CMSV) of a linear operator to measure the stability of low-rank matrix reconstruction. By employing advanced tools from geometrical functional analysis and empirical processes, we show that a large class of random linear operators have $\ell_*$-CMSVs bounded away from zero. We also derive a fixed point characterization of the $\ell_*$-CMSV.

Several works in the literature also address the problem of low-rank matrix reconstruction. Recht \emph{et.al.} study the recovery of $X$ in model \eqref{intro:model} in the noiseless setting \cite{Recht2010Nuclear}. The matrix restricted isometry property (mRIP) is shown to guarantee exact recovery of $X$ subject to the measurement constraint $\A(X) = \y$. Cand\'{e}s \emph{et.al.} consider the noisy problem and analyze the reconstruction performance of several convex relaxation algorithms \cite{candes2009lowrank}. The techniques used in this paper for deriving the error bounds in terms of $\ell_*$-CMSV draw ideas from \cite{candes2009lowrank}. In both works \cite{Recht2010Nuclear} and \cite{candes2009lowrank}, several important random measurement ensembles are shown to have the matrix restricted isometry constant (mRIC) close to zero for reasonably large $m$. Our procedures for establishing the parallel results for the $\ell_*$-CMSV are significantly different from those in \cite{Recht2010Nuclear} and \cite{candes2009lowrank}. In particular, for correlated Gaussian operators, we show that the mRIC might fail with high probability while the $\ell_*$-CMSV is still well controlled.

The $\ell_*$-CMSV has several advantages over the mRIC in stability analysis of low-rank matrix analysis. First, the error bounds involving $\ell_*$-CMSV have more transparent relationships with the Signal-to-Noise-Ratio. For example, consider the matrix Basis Pursuit algorithm, if we multiply the measurement operator $\A$ by a positive constant, the $\ell_*$-CMSV will scale by the same constant and the error bound for the matrix Basis Pursuit will scale inverse proportionally, while the mRIC and associated error bounds have more complex scaling properties. Second, the $\ell_*$-CMSV shows clear relations of low-rank matrix recovery with certain geometric properties of the nuclear ball, such as its mean width, Gaussian width, and the diameter of its sections. In addition, the derivation of the $\ell_*$-CMSV bounds is less complicated and the resulting bounds have more concise forms. Last but not least, as shown by our probabilistic analysis for correlated Gaussian operators, the mRIC might fail while the $\ell_*$-CMSV is still meaningful.

The paper is organized as follows. Section \ref{sec:model} introduces notation, the measurement model, three convex relaxation based recovery algorithms, and the definition and properties of the mRIC. Section \ref{sec:errorbound} is devoted to deriving error bounds in terms of the $\ell_*$-CMSV for three convex relaxation algorithms. In Section \ref{sec:random} we analyze the $\ell_*$-CMSV for isotropic and subgaussian measurement operators, and in Section \ref{sec:fixedpoint} we provide a fixed point characterization of the $\ell_*$-CMSV. The paper is concluded in Section \ref{sec:conclusions}.

\section{Notation, Measurement Model, Reconstruction Algorithms, and Matrix Restricted Isometry Constant}\label{sec:model}
\subsection{Notation}\label{subsec:notations}
We use bold lower case letters such as $\x, \y, \z$ to denote vectors whose $i$th components are represented by corresponding lower case letters $x_i$,  $y_i$, and $z_i$, respectively. Subscripted bold lower case letters, e.g., $\x_i$, are reserved for vectors with subscript $i$. Matrices are denoted by upper case letters such as $A$, $X$, $Z$.

The $\ell_p$ norm $\|\cdot\|_p$ of $\x = [x_1,\ldots,x_m]^T  \in \R^m$ is defined as
\begin{eqnarray}
\|\x\|_p = (\sum_{k\leq m}|x_k|^p)^{1/p} \ \ \text{for}\ \ 1\leq p < \infty
\end{eqnarray}
and
\begin{eqnarray}
  \|\x\|_\infty = \max_{k \leq m}|x_k|.
\end{eqnarray}

Suppose $X = [\x_1\ \x_2\ \ldots\ \x_{n_2}] \in \R^{n_1\times n_2}$ is a matrix. Define the Frobenius norm of $X$ as $\| X\|_{\rmF}=(\sum_{i,j} |X_{ij}|^2)^{1/2} = (\sum_{i} \sigma_i^2(X))^{1/2}$, the nuclear norm as $\| X\|_*=\sum_{i} {\sigma }_i(X)$, and the operator norm as $\|X\| = \max\{\sigma_i(X)\}$, where ${\sigma }_i(X)$ is the $i$th singular value of $X$ in descending order. The rank and trace of $X$ are denoted by $\rank(X)$ and $\tr(X)$, respectively. The inner product of two matrices $X_1, X_2 \in \R^{n_1 \times n_2}$ is defined as $\left <X_1, X_2\right> = \tr(X_1^TX_2)$. We use $\otimes$ to denote the Kronecker product, and $\vec(\cdot)$ to denote the vectorization of a matrix. A useful identity is $(B^T\otimes A) \vec(X) = \vec(AXB)$.




For any linear operator $\A: \R^{n_1\times n_2} \mapsto \R^m$, its adjoint operator $\A^*: \R^m \mapsto \R^{n_1 \times n_2}$ is defined by the relation
\begin{eqnarray}
   \left<\A(X), \z\right> = \left<X, \A^*(\z)\right>, \forall X \in \R^{n_1\times n_2}, \z \in \R^m.
\end{eqnarray}
A linear operator $\A: \R^{n_1\times n_2} \mapsto \R^m$ can be represented by $m$ matrices $\{A^1, A^2, \ldots, A^m\} \subset \R^{n_1\times n_2}$ such that $\A(X) = \left[\left<A^1,X\right>, \ldots, \left<A^m,X\right>\right]^T$, or by a big matrix $A\in
\R^{m\times n_1n_2}$ whose $k$th row is $\vec(A^k)$ such that $\A(X) = A\vec(X)$. The adjoint operator is given by
\begin{eqnarray}
  \A^*(\z) &=& \sum_{k=1}^m z_k A^k \in \R^{n_1\times n_2}.
\end{eqnarray}

Gaussian distribution with mean $\mu$ and variance $\sigma^2$ is denoted by $\N(\mu, \sigma^2)$. This notation is also generalized to Gaussian random vectors and matrix variate Gaussian distributions \cite{Gupta1999Matrix}.


\subsection{The Measurement Model}\label{subsec:measmodel}
Throughout the paper, we will assume $n_1 \leq n_2$. Suppose we have a matrix $X \in \R^{n_1\times n_2}$ with $\rank(X) = r \ll n_1$. We observe $X$ through the following linear model:
\begin{eqnarray}\label{model}
  \y &=& \A(X) + \w,
\end{eqnarray}
where $\A: \R^{n_1 \times n_2} \mapsto \R^m$ is a linear operator and $\w \in \R^m$ is noise. Here $m$ is much less than $n_1n_2$.

A fundamental problem pertaining to model \eqref{model} is to reconstruct the low-rank matrix $X$ from the measurement $\y$ by exploiting the low-rank property of $X$, and the stability of the reconstruction with respect to noise. For any reconstruction algorithm, we denote the estimate of $X$ as $\hX$, and the error matrix $H \df \hX - X$. In this paper, the stability problem aims to bound $\|H\|_\rmF$ in terms of $m, n_1, n_2, r$, the linear operator $\A$, and the noise level.

\subsection{Reconstruction Algorithms}\label{subsec:algorithms}
We consider three low-rank matrix recovery algorithms based on convex relaxation: the matrix Basis Pursuit (mBP), the matrix Dantzig selector (mDS), and the matrix LASSO estimator (mLASSO). 

The mBP algorithm \cite{Recht2010Nuclear, candes2009lowrank} minimizes the nuclear norm subject to bounded noise constraint:
\begin{eqnarray}\label{mbp}
\mathrm{mBP:}\ \min_{Z \in \R^{n_1\times n_2}}\|Z\|_* \text{\ s.t.\ } \|\y - \A(Z)\|_2 \leq \varepsilon.
\end{eqnarray}
The mDS \cite{candes2009lowrank} reconstructs a low-rank matrix when its linear measurements are corrupted by unbounded noise. Its estimate for $X$ is the solution to the nuclear norm regularization problem:
\begin{eqnarray}\label{mds}
  \mathrm{mDS:}\min_{Z \in \R^{n_1\times n_2}}\|Z\|_* \text{\ s.t.\ } \|\A^* (\y - \A(Z)) \| \leq \mu.
\end{eqnarray}
The mLASSO solves the following optimization problem \cite{ma2009bregman, candes2009lowrank}:
\begin{eqnarray}\label{mlasso}
  \text{mLASSO:\ \ \ }\min_{Z \in \R^{n_1 \times n_2}} \frac{1}{2}\|\y - \A(Z)\|_2^2 + \mu \|Z\|_*.
\end{eqnarray}
All three optimization problems can be solved using semidefinite programs.

\subsection{Matrix Restricted Isometry Constant}\label{subsec:mric}
The reconstruction performance of the mBP, the mDS, and the mLASSO depends on the incoherence of the linear operator $\A$. A popular measure of incoherence is mRIC defined below \cite{Recht2010Nuclear, candes2009lowrank}:
\begin{defi}\label{def:mric}
For each integer $r \in \{1,\ldots, n_1\}$, the matrix restricted isometry constant (mRIC) $\delta_r$ of a linear operator $\A: \R^{n_1 \times n_2} \mapsto \R^m$ is defined as the smallest $\delta > 0$ such that
\begin{eqnarray}
1-\delta \leq \frac{\|\A(X)\|_2^2}{\|X\|_\rmF^2} \leq 1+\delta
\end{eqnarray}
holds for arbitrary non-zero matrix $X$ of rank at most $r$.
\end{defi}
A linear operator $\A$ with a small $\delta_r$ roughly means that $\A$ is nearly an isometry when restricted onto all matrices with rank at most $r$. 
We cite stability results on the mBP, the mDS and the mLASSO, which are expressed in terms of the mRIC. Assume $X$ is of rank $r$ and $\hX$ is its estimate given by any of the three algorithms; then we have the following:
\begin{enumerate}
  \item mBP\cite{candes2009lowrank}: Suppose that $\delta_{4r} < \sqrt{2}-1$ and $\|\w\|_2 \leq \varepsilon$. The solution to the mBP \eqref{mbp} satisfies
  \begin{eqnarray}\label{bp_rip_bd}
\|\hX-X\|_\rmF \leq \frac{4\sqrt{1+\delta_{4r}}}{1-(1+\sqrt{2})\delta_{4r}} \varepsilon.
\end{eqnarray}
  \item mDS\cite{candes2009lowrank}: If $\delta_{4r} < \sqrt{2}-1$ and $\|\A^*(\w)\| \leq \mu$, then
      \begin{eqnarray}\label{ds_rip_bd}
        \|\hX-X\|_\rmF \leq \frac{16 \sqrt{r}}{1-(\sqrt{2}+1)\delta_{4r}} \mu.
      \end{eqnarray}
  \item mLASSO \cite{candes2009lowrank}: If $\delta_{4r} < (3\sqrt{2}-1)/17$ and $\|\A^*(\w)\| \leq \mu/2$, then the solution to the mLASSO \eqref{mlasso} satisfies
      \begin{eqnarray}\label{lasso_rip_bd}
        \|\hX - X\|_\rmF \leq C\delta_{4r} \sqrt{r} \mu,
      \end{eqnarray}
      for some numerical constant $C$.
\end{enumerate}


\section{Recovery Error Bounds}\label{sec:errorbound}
In this section, we derive bounds on the recovery errors of the mBP, the mDS, and mLASSO. We first characterize the recovery errors by showing that the effective ranks of the error matrices are small.
\subsection{Error Characteristics}\label{sec:csv}
We introduce a quantity that continuously extends the concept of rank for a given matrix $X$.
\begin{defi}
The $\ell_*$-rank of a non-zero matrix $X \in \R^{n_1 \times n_2}$ is defined as
\begin{eqnarray}
  \tau(X) = \frac{\|X\|_*^2}{\|X\|_\rmF^2}.
 \end{eqnarray}
\end{defi}

The function $\tau(X)$ is indeed a measure of the effective rank. To see this, suppose $\rank(X) = r$; then Cauchy-Schwarz inequality implies that
\begin{eqnarray}\label{sp_bd}
  \tau(X)  \leq r,
\end{eqnarray}
and we have equality if and only if all non-zero singular values of $X$ are equal. Therefore, the more non-zero singular values $X$ has and the more evenly the magnitudes of these non-zero singular values are distributed, the larger $\tau(X)$. In particular, if $X$ is of rank $1$, then $\tau(X) = 1$; if $X$ is of full rank $n_1$ with all singular values having the same magnitudes, then $\tau(X) = n_1$. However, if $X$ has $n_1$ non-zero singular values but their magnitudes are spread in a wide range, then its $\ell_*$-rank might be very small.

The following proposition, whose proof is given in Appendix \ref{app:pf:pro:errorcharacteristics}, shows that the error matrices have small $\ell_*$-rank:
\begin{pro}\label{pro:errorcharacteristics}
Suppose $X$ in \eqref{model} is of rank $r$ and the noise $\w$ satisfies $\|\w\|_2 \leq \varepsilon$, $\|\A^*(\w)\| \leq \mu$, and $\|\A^*(\w)\| \leq \kappa \mu, \kappa \in (0,1)$, for the mBP, the mDS, and the mLASSO, respectively. Then the error matrix $H = \hX - X$ for any of the three recovery algorithms \eqref{mbp}, \eqref{mds}, and \eqref{mlasso} satisfies
\begin{eqnarray}
\tau(H) \leq cr 
\end{eqnarray}
where $ c = 8$ for the mBP and the mDS, and $c = 8/(1-\kappa)^2$ for the mLASSO.
\end{pro}

\subsection{$\ell_*$-CMSV and Error Bounds}
The reconstruction performance of the recovery algorithms should depend on the invertibility of the linear operator $\A$. Proposition \ref{pro:errorcharacteristics} indicates that we could restrict ourselves to the set $\{X \in \R^{n_1\times n_2}: \tau(X) \leq cr \}$ when quantifying the invertability of $\A$. 
\begin{defi}\label{def:lstarcmsv}
For any $\tau \in [1, n_1]$ and any linear operator $\A: \R^{n_1\times n_2} \mapsto \R^m$, define the $\ell_*$-constrained minimal singular value (abbreviated as $\ell_*$-CMSV) of $\A$ by
\begin{eqnarray}
\rho_\tau(\A) &:=& \inf_{X \neq 0,\ \tau(X) \leq \tau } \frac{\|\A(X)\|_2}{\|X\|_\rmF}.
\end{eqnarray}
\end{defi}

As pointed out by one reviewer, one difference between the $\ell_*$-CMSV and the mRIC is that the $\ell_*$-CMSV does not require upper bounds on the restricted eigenvalues of the operator $\A$. This is known to be true in the vector case (\emph{i.e.}, sparsity recovery) and is established using the notions such as \emph{Restricted Eigenvalues} \cite{bickel2009simultaneous} and \emph{$m$-Sparse Minimal Eigenvalues} \cite{meinshausen2009lassotype}. This paper shares the common observation with previous work on the vector case that the reconstruction performance of the recovery algorithms should depend on the invertibility of the measurement matrix or operator when restricted to the error set, which is usually much smaller than the the signal's ambient space. We use the $\ell_*$-rank to differentiate the error set and to define the invertibility of $\A$. Probability analysis in Section IV shows that at least for isotropic and subgaussian operators, the $\ell_*$-rank characterization is as good as the null space property characterization \cite{recht2011matrixNSP}. We also establish that, for correlated Gaussian operators, the mRIC might not be valid with high probability, even when the $\ell_*$-CMSV is still bounded away from zero.

Now we present bounds on the error matrices for the mBP, the mDS, and the mLASSO. The proof is given in Appendix \ref{app:pf:thm:errorbound}.
\begin{thm}\label{thm:errorbound}
Under the assumption of Proposition \ref{pro:errorcharacteristics}, we have
\begin{eqnarray}
  \|\hX - X\|_\rmF \leq \frac{2\varepsilon}{\rho_{8r}(\A)}
\end{eqnarray}
for the mBP,
\begin{eqnarray}
  \|\hX-X\|_\rmF \leq \frac{4\sqrt{2r}}{\rho_{8r}^2(\A)}\mu
\end{eqnarray}
for the mDS, and
\begin{eqnarray}\label{bd:mlasso}
  \|\hX-X\|_\rmF \leq \frac{1+\kappa}{1-\kappa}\frac{2\sqrt{2r}}{\rho_{\frac{8r}{(1-\kappa)^2}}^2(\A)} \mu
\end{eqnarray}
for the mLASSO.
\end{thm}

Compared with the error bounds \eqref{bp_rip_bd}, \eqref{ds_rip_bd}, and \eqref{lasso_rip_bd}, the bounds given in Theorem \ref{thm:errorbound} are simpler and their derivations are easier. When the noise levels are zero, namely, $\varepsilon = 0$ and $\mu = 0$, roughly speaking all three nuclear norm minimization algorithms reduce to 
\begin{eqnarray}\label{eqn:noise_free}
\min_{Z \in \R^{n_1\times n_2}} \|Z\|_* \text{\ subject to \ }  \y = \A(Z). 
\end{eqnarray}
According to Theorem \ref{thm:errorbound}, if $\rho_{8r}(\A) > 0$, then we get exact recovery in the noise-free case. Therefore, $\rho_{8r} > 0$ is a sufficient condition for exact low-rank matrix recovery using nuclear norm minimization. 

We observe that $\rho_{8r}(\A) > 0$ is equivalent to 
\begin{eqnarray}
r < \frac{1}{8} \min\{\tau(Z): \A(Z) = 0\}, \text{\ or}\nn\\
\frac{1}{2\sqrt{2r}} > \max \{\|Z\|_\rmF: \A(Z) = 0, \|Z\|_* \leq 1\}. \label{eqn:diam}
\end{eqnarray}
We note that the right hand side of \eqref{eqn:diam} is the diamdeter of the set $B_*^{n_1\times n_2} \bigcap \mathrm{null}(\A)$, where $B_*^{n_1\times n_2}$ is the unit nuclear ball and $\mathrm{null}(\A)$ is the null space of the operator $\A$. If the null space $\mathrm{null}(\A)$ is chosen uniformly according to the Haar measure on the Grassmanian $\mathcal{G}_{n_1n_2, n_1n_2-m}$ of $(n_1n_2-m)$-dimensional subspaces of $\R^{n_1\times n_2}$ (\emph{e.g.} when the entries of $\{A_i\}_{i=1}^m$ follow \emph{i.i.d.} Gaussian with zero mean and unit variance), then the low $M^*$ estimate \cite{Boroczky2010meanwidth} implies that
\begin{eqnarray}
&&\mathrm{diam}(B_*^{n_1\times n_2} \bigcap \mathrm{null}(\A))\nn\\
 &:=& \max_{Z: Z \in B_*^{n_1\times n_2} \bigcap\mathrm{null}(\A)}  \|Z\|_\rmF\nn\\
& \leq& c \sqrt{\frac{n_1n_2}{m}} M^*(B_*^{n_1\times n_2}) \nn\\
&:=& c \sqrt{\frac{n_1n_2}{m}} \int_{\mathbb{S}^{n_1n_2 - 1}}\|Z\| \mathrm{d}\sigma(Z)
\end{eqnarray}
with probability at least $1 - e^{-m}$. Here $\mathbb{S}^{n_1n_2-1}$ is the unit Euclidean sphere in $\R^{n_1\times n_2}$, $\sigma(\cdot)$ is the Haar measure on $\mathbb{S}^{n_1n_2-1}$, and $c$ is a numerical constant.

According to Poincar\'e's lemma \cite{stroock1993probability}, the uniform measures on $n$-dimensional spheres with radius $\sqrt{n}$ approximate Gaussian measures. As a consequence, we have
\begin{eqnarray}
&&M^*(B_*^{n_1\times n_2}) \nn\\
&=& \frac{1}{\sqrt{n_1n_2}}\int_{\sqrt{n_1n_2}\mathbb{S}^{n_1n_2 - 1}}\|Z\| \mathrm{d}\sqrt{n_1n_2}\sigma(Z)\nn\\
& \sim & \frac{1}{\sqrt{n_1n_2}}\E \|Z\| \leq 2 \sqrt{\frac{n_2}{n_1n_2}}.
\end{eqnarray} 
Here $\E$ is taken with respect to the canonical Gaussian measure in $\R^{n_1n_2}$, and the last inequality, which gives an upper bound on the expected largest singular value of a rectangular Gaussian matrix, is due to Slepian's lemma \cite{davidson2001local, vershyninlecture}. Therefore, we obtain 
\begin{eqnarray}
\mathrm{diam}(B_*^{n_1\times n_2}\bigcap \mathrm{null}(\A)) \leq c \sqrt{\frac{n_2}{m}}
\end{eqnarray}
with high probability. Combining with the sufficient condition \eqref{eqn:diam}, we obtain the following corollary:
\begin{cor}\label{cor:uniform}
If the null space of $\A$ follows uniform distribtion on all subspaces of dimension $n_1n_2 - m$ and 
\begin{eqnarray}
m \geq cn_2 r,
\end{eqnarray}
then with probability greater than $1 - e^{-m}$ we can recover any matrix $X$ of rank less than $r$. 
\end{cor}

In the next section, we will directly analyze the proabilistic behavior of $\rho_\tau(\A)$ and obtain Corollary \ref{cor:uniform} as a consequence. 

\section{Probabilistic Analysis}\label{sec:random}
This section is devoted to analyzing the properties of the $\ell_*$-CMSVs for several important random operator ensembles. Although the bounds in Theorem \ref{thm:errorbound} have concise forms, they are useless if the quantity involved, $\rho_\tau$, is zero or approaches zero for most matrices as $n_1, n_2, m, r$ vary in a reasonable manner. We show that for a large class of random linear operators, including both isotropic and subgaussian operators and correlated Gaussian operators, the $\ell_*$-CMSVs are bounded away from zero with high probability.

\subsection{Isotropic and subgaussian operators}
We begin by defining the isotropic and subgaussian ensemble, after introducing some notations. For a scalar random variable $a$, the Orlicz $\psi_2$ norm \cite[Section 4.1, page 92]{ledoux1991probability} is defined as
\begin{eqnarray}
  \|a\|_{\psi_2} = \inf \left\{t > 0: \E \exp\left(\frac{|a|^2}{t^2}\right) \leq 2\right\}.
\end{eqnarray}
Markov's inequality immediately gives that $a$ with finite $\|a\|_{\psi_2}$ has a subgaussian tail:
\begin{eqnarray}
\Pr(|a| \geq t ) \leq 2 \exp(-ct^2/\|a\|_{\psi_2}).
\end{eqnarray}
The converse is also true, \emph{i.e.,} if $a$ has subgaussian tail $\exp(-t^2/K^2)$, then $\|a\|_{\psi_2} \leq c K$. A random vector $\a \in \R^n$ is called \emph{isotropic and subgaussian with constant $L$} if $\E|\left<\a, \u\right>|^2 = \|\u\|_2^2$ and $\|\left<\a, \u\right>\|_{\psi_2} \leq L \|\u\|_2$ hold for any $\u \in \R^n$.

Recall that a linear operator $\A: \R^{n_1\times n_2}\rightarrow \R^m$ can be represented by a collection of matrices $\{A^1, \ldots, A^m\}$. Based on this representation of $\A$, we have the following definition of isotropic and subgaussian operators:
\begin{defi}\label{def:subgaussianoperator}
Suppose $\A: \R^{n_1\times n_2}\rightarrow \R^m$ is a linear operator with corresponding matrix representation $\{A^i\}_{i=1}^m$. We say $\A$ is from the isotropic and subgaussian ensemble if $\{A^i\}_{i=1}^m$ are independent isotropic and subgaussian vector with constant $L$, where $L$ is a numerical constant independent of $n_1, n_2$.
\end{defi}

Isotropic and subgaussian operators include operators with {i.i.d} centered subgaussian entries of unit variance (Gaussian and Bernoulli entries in particular) as well as operators whose matrices $A_i$ ($\vec(A_i)$, more precisely) are independent copies of random vectors distributed according to the normalized volume measure of unit balls of $(\R^{n_1n_2}, \|\cdot\|_p)$ for $2\leq p \leq \infty$.

An important concept in studying empirical processes of isotropic and subgaussian random vectors is the Gaussian width defined below:

\begin{defi}
Let $\H \subset \R^n$ and the components of $\g$ follow i.i.d. Gaussian with mean zero and variance one, i.e., $\g \sim \N(0,\I_n)$. Denote by $w(\H) = \E\ \sup_{\u \in \H} \left<\g, \u\right>$.
\end{defi}

With these preparations, we combine \cite[Theorem D]{mendelson2007subgaussian} and the discussion below it to present:
\begin{thm}\cite[Theorem D]{mendelson2007subgaussian}\label{thm:empirical}
Let $\{\a, \a_i, i = 1,\ldots,m\} \subset \R^n$ be \emph{i.i.d.} isotropic and subgaussian random vectors, $\H$ be a subset of the unit sphere of $\R^n$, and $\sF = \{f_{\u}(\cdot) = \left<\u, \cdot\right>: \u \in \H\}$. Suppose $\mathrm{diam}(\sF, \|\cdot\|_{\psi_2}) = \max_{f, g\in \sF} \|f-g\|_{\psi_2} = \alpha$. Then there exist absolute constants $c_1, c_2, c_3$ such that for any $\epsilon > 0$ and $m \geq 1$ satisfying
\begin{eqnarray}
  m \geq c_1 \frac{\alpha^2 w^2(\H)}{\epsilon^2},
\end{eqnarray}
with probability at least $1 - \exp(-c_2\epsilon^2 m/\alpha^4)$,
\begin{eqnarray}
  \sup_{f\in \sF}\left|\frac{1}{m} \sum_{k=1}^m f^2(\a_k) - \E f^2(\a)\right| \leq \epsilon.
\end{eqnarray}
Furthermore, if $\sF$ is symmetric, we have
\begin{eqnarray}
\hskip -1cm &&\E \sup_{f \in \sF} \left|\frac{1}{m} \sum_{k=1}^m f^2(\a_k) - \E f^2(\a)\right| \nn\\
\hskip -1cm &&\leq c_3 \max\left\{\alpha \frac{w(\H)}{\sqrt{m}}, \frac{w^2(\H)}{m}\right\}.
\end{eqnarray}
\end{thm}

Using Theorem \ref{thm:empirical}, we show that for any isotropic and subgaussian operator $\sqrt{m}\A$ the typical value of $\rho_\tau(\A)$ concentrates around $1$ for relatively large $m$ (but $\ll n_1n_2$). More precisely, we have the following theorem:
\begin{thm}\label{thm:matrix_randomcmsv}
Let $\sqrt{m}\A$ be an isotropic and subgaussian operator with some numerical constant $L$. Then there exist absolute constants $c_1, c_2$ depending on $L$ only such that for any $\epsilon > 0$ and $m \geq 1$ satisfying
\begin{eqnarray}
  m \geq c_1 \frac{\tau  n_2}{\epsilon^2},
\end{eqnarray}
we have
\begin{eqnarray}
\E|\rho_\tau^2(\A)-1| \leq \epsilon
\end{eqnarray}
and
\begin{eqnarray}
\Pr\left\{|\rho_\tau^2(\A)-1|\leq \epsilon\right\} \geq 1 - \exp(-c_2 \epsilon^2 m).
\end{eqnarray}
\end{thm}

\begin{proof}[Proof of Theorem \ref{thm:matrix_randomcmsv}]
Since the linear operator $\A$ is generated in a way such that $\E \left<\sqrt{m}A^k, X\right>^2 = \|X\|_{\rmF}^2$ for any $X \in \R^{n_1\times n_2}$, we have $|\rho_\tau^2(\A) - 1| < 1 - \epsilon$ is a consequence of
\begin{eqnarray}\label{eqn:supform}
  &&\sup_{X \in \H_\tau}\left|\A(X)^T\A(X) - 1\right|\nn\\
   &=&  \sup_{X \in \H_\tau}\left|\frac{1}{m}\sum_{k=1}^m \left<\sqrt{m}A^k, X\right>^2 - 1\right| \leq \epsilon.
\end{eqnarray}
Here the set
\begin{eqnarray}
\H_\tau = \{X \in \R^{n_1\times n_2}: \|X\|_{\rmF} = 1,\ \|X\|_*^2 \leq \tau \}.
\end{eqnarray}
As usual, the operator $\A$ is represented by a collection of matrices $\{A^1, \ldots, A^m\}$. We define a class of functions parameterized by $X$ as $\sF_\tau := \{f_{X}(\cdot) = \left<X, \cdot\right>: X \in \H_\tau\}$.

It remains to compute $w(\H_\tau)$ as follows
\begin{eqnarray}
  w(\H_\tau) &=& \E \sup_{X \in \H_\tau} \left<G, X\right> \nn\\
  &\leq& c\  \|X\|_* \ \E\ \|G\|_2 \nn\\
  &\leq& c\ \sqrt{\tau} \sqrt{n_2},
\end{eqnarray}
where $G$ is a Gaussian matrix with \emph{i.i.d.} entries from $\N(0,1)$. Again we have used an upper bound on the expected largest singular value of a rectangular Gaussian matrix due to Slepian's lemma \cite[Chapter 3.1]{ledoux1991probability}. As a consequence, the conclusions of Theorem \ref{thm:matrix_randomcmsv} hold.
\end{proof}

If we take $\epsilon = 1/2$ and $\tau = 8r$ in Theorem \ref{thm:matrix_randomcmsv}, we get Corollary \ref{cor:uniform} as a consequence. The bound $m = \Omega(r n_2)$ is the same as the one obtained for the mRIC. Thus, the $\ell_*$-CMSV is as good as the mRIC for isotropic and subgaussian operators.

\subsection{Correlated Gaussian operators}
In this subsection, we consider Gaussian measurement operators with a
correlation structure. Correlated sensing matrices are considered in \cite{Raskutti2010correlated} and \cite{Rudelson2011anisotropic}
in the context of compressive sensing. For low-rank matrix recovery,
correlated measurement operators are potentially useful for multivariate
regression and vector autoregressive processes \cite{Negahban2009near}.

Suppose that the entries of $G^k \in \mathbbm{R}^{n_1 \times n_2}, k = 1,
\ldots, m$ follow i.i.d. Gaussian distribution $\mathcal{N} \left( 0,
\frac{1}{m} \right)$, and $\Sigma_1 \in \mathbbm{R}^{n_1 \times n_1}$ and
$\Sigma_2 \in \mathbbm{R}^{n_2 \times n_2}$ are positive semidefinite
matrices. Then $\Sigma_1^{1 / 2} G^k \Sigma_2^{1 / 2}, k = 1, \ldots, m$
follow i.i.d. matrix variate Gaussian distribution $\mathcal{N} \left( 0,
\frac{1}{m} \Sigma_2 \otimes \Sigma_1 \right)$, namely, \ $\tmop{vec} \left(
\Sigma_1^{1 / 2} G^k \Sigma_2^{1 / 2} \right) \sim \mathcal{N} \left( 0,
\frac{1}{m} \Sigma_2 \otimes \Sigma_1 \right)$\cite{Gupta1999Matrix}. Here $\Sigma^{1 / 2}$
denotes the matrix square root of a positive semidefiite matrix $\Sigma$. We
call the linear operator $\mathcal{A}_{\Sigma_1, \Sigma_2}$ represented by $\left\{
\Sigma_1^{1 / 2} G^k \Sigma_2^{1 / 2}, k = 1, \ldots, m \right\}$ a correlated
Gaussian measurement operator.

The following theorem shows that $\rho_{\tau} \left( \mathcal{A}_{\Sigma_1,
\Sigma_2} \right)$ is controlled by $\rho_{\tau} \left( \Sigma_2^{1 / 2}
\otimes \Sigma_1^{1 / 2} \right)$ with high probability.

\begin{thm}
  \label{thm:correlated:Gaussian}Suppose $\rho_{\tau} \left( \Sigma_2^{1 / 2}
  \otimes \Sigma_1^{1 / 2} \right) > 0$. Then there exist universal positive
  constant $c, c_1, c_2$ such that if
  \begin{eqnarray}
    m & \geq & c\ \frac{\tmop{trace} \left( \Sigma_1 \right) + \tmop{trace}
    \left( \Sigma_2 \right)}{\rho_{\tau}^2 \left( \Sigma_2^{1 / 2} \otimes
    \Sigma_1^{1 / 2} \right)} \tau,
  \end{eqnarray}
  then the linear operator $\mathcal{A}_{\Sigma_1, \Sigma_2}$ has
  $\ell_{\ast}$-CMSV
  \begin{eqnarray}
    \rho_{\tau} \left( \mathcal{A}_{\Sigma_1, \Sigma_2} \right) & \geq &
    \frac{1}{8} \rho_{\tau} \left( \Sigma_2^{1 / 2} \otimes \Sigma_1^{1 / 2}
    \right)
  \end{eqnarray}
  with probability at least $1 - c_1 \exp \left( - c_2 m \right)$.
\end{thm}

\begin{remark}
  When $\Sigma_1 = I_{n_1}$ and $\Sigma_2 = I_{n_2}$, clearly we have
  $\rho_{\tau} \left( \Sigma_2^{1 / 2} \otimes \Sigma_1^{1 / 2} \right) =
  \rho_{\tau} \left( I_{n_1 n_2} \right) = 1$, and $\tmop{trace} \left(
  \Sigma_1 \right) + \tmop{trace} \left( \Sigma_2 \right) = n_1 + n_2$. So
  Theorem \ref{thm:correlated:Gaussian} is consistent with Theorem
  \ref{thm:matrix_randomcmsv} applied to Gaussian measurement operators.
\end{remark}

\begin{remark}
  In this remark, we provide an example where the $\ell_{\ast}$-CMSV is well
  controlled while the mRIP fails with high probability. Refere to \cite{Raskutti2010correlated} for more
  examples constructed in the similar compressive sensing setting. For
  simplicity we set $n_1 = n_2 = n$. Consider $\Sigma_1 = \Sigma_2 = \Sigma =
  \left( 1 - a \right) I_n + a\tmmathbf{1}\tmmathbf{1}^T$ for some fixed $a \in
  \left( 0, 1 \right)$. Here $\tmmathbf{1} \in \R^n$ is the vector with all ones. Clearly, we have
  $\tmop{trace} \left( \Sigma \right) = n$. The inequality $\rho_{\tau}^2
  \left( \Sigma^{1 / 2} \otimes \Sigma^{1 / 2} \right) \geq \lambda_{\min}
  \left( \Sigma \otimes \Sigma \right) = \lambda_{\min}^2 \left( \Sigma
  \right) = \left( 1 - a \right)^2$ together with Theorem
  \ref{thm:correlated:Gaussian} imply that $\begin{array}{lll}
    \rho_{\tau} \left( \mathcal{A}_{\Sigma_1, \Sigma_2} \right) & \geq &
    \frac{1}{8} \left( 1 - a \right)
  \end{array}$with high probability as long as $m \geq c \frac{n \tau}{\left(
  1 - a \right)^2}$.
  
  However, $\Sigma^{1 / 2} \otimes \Sigma^{1 / 2}$ does not satisfy the mRIC
  when $n$ is large. To see this, assume that the eigen-decomposition of
  $\Sigma$ is $\sum_{j = 1}^n \sigma_j \tmmathbf{u}_j \tmmathbf{u}_j^T$ and
  observe that when $X = \sum_{i = 1}^r \lambda_i \tmmathbf{u}_{j_i}
  \tmmathbf{u}_{j_i}^T$ we have
  \begin{eqnarray}
    &&\frac{\left\| \left( \Sigma^{1 / 2} \otimes \Sigma^{1 / 2} \right)
    \tmop{vec} \left( X \right) \right\|_2^2}{\left\| X \right\|_F^2}\nonumber\\
    & = &
    \frac{\left\| \Sigma^{1 / 2} X \Sigma^{1 / 2} \right\|_F^2}{\left\| X
    \right\|_F^2} = \frac{\sum_{i = 1}^r \sigma_{j_i}^2 \lambda_i^2}{\sum_{i =
    1}^r \lambda_i^2} .
  \end{eqnarray}
  Taking supremum and infimum leads to
  \begin{eqnarray}
    && \sup_{X : \tmop{rank} \left( X \right) \leq r} \frac{\left\| \left(
    \Sigma^{1 / 2} \otimes \Sigma^{1 / 2} \right) \tmop{vec} \left( X \right)
    \right\|_2^2}{\left\| X \right\|_F^2}\nonumber \\
    & \geq & \sup_{\lambda i}
    \frac{\sum_{i = 1}^r \sigma_{j_i}^2 \lambda_i^2}{\sum_{i = 1}^r
    \lambda_i^2} = \max \left\{ \sigma_i^2 \right\}\nn\\
    & = & \left( \left( 1 - a
    \right) + n a \right)^2\\
   && \inf_{X : \tmop{rank} \left( X \right) \leq r} \frac{\left\| \left(
    \Sigma^{1 / 2} \otimes \Sigma^{1 / 2} \right) \tmop{vec} \left( X \right)
    \right\|_2^2}{\left\| X \right\|_F^2}\nn
    \\ & \leq & \inf_{\lambda i}
    \frac{\sum_{i = 1}^r \sigma_{j_i}^2 \lambda_i^2}{\sum_{i = 1}^r
    \lambda_i^2} = \min \left\{ \sigma_i^2 \right\}\nn\\
    & =& \left( 1 - a \right)^2 .
  \end{eqnarray}
  Therefore, independent of the rank parameter, $\Sigma^{1 / 2} \otimes
  \Sigma^{1 / 2}$ does not satisfy the mRIC when $n$ is large. A large
  deviation argument similar to the one given in \cite{Raskutti2010correlated} shows that the same
  statment is true with high probability for $\mathcal{A}_{\Sigma_1,
  \Sigma_2}$. Hence, the mRIP might be violated with high probability for
  correlated Gaussian operators while the $\ell_{\ast}$-CMSV is still well
  controlled. Roughly speaking, the problem with mRIC is that it involves the
  maximal eigenvalue while only the minimal eigenvalue is essential for
  low-rank matrix recovery.
\end{remark}

The proof of Theorem \ref{thm:correlated:Gaussian} is based on the following
proposition:

\begin{pro}
  \label{pro:gaussian:estimate}
  If each matrix $A^k$ associated with
  $\mathcal{A}_{\Sigma_1, \Sigma_2}$ are i.i.d. random matrices following 
  $\mathcal{N} \left( 0, \frac{1}{m}
  \Sigma_2 \otimes \Sigma_1 \right)$, then there exist univeral constant $c,
  c_1$ such that for all $ X
    \in \mathbbm{R}^{n_1 \times n_2}$
  \begin{eqnarray}
  \hspace{-1cm}&&\left\| \mathcal{A}_{\Sigma_1, \Sigma_2} \left( X \right) \right\|_2
    \geq \frac{1}{4} \left\| \Sigma_1^{1 / 2} X \Sigma_2^{1 / 2} \right\|_F\nn\\
     \hspace{-1cm}&&  \ \ \ \ \ \ - 3 \frac{\sqrt{\tmop{trace} \left( \Sigma_1 \right)} + \sqrt{\tmop{trace}
    \left( \Sigma_2 \right)}}{\sqrt{m}} \left\| X \right\|_{\ast},\label{eqn:gaussian:estimate}
  \end{eqnarray}
  with probability at least $1 - c_1 \exp \left( - c m \right)$.
\end{pro}

\begin{proof}[Proof of Theorem \ref{thm:correlated:Gaussian}]
  By the definition of $\ell_{\ast}$-CMSV, for any $X$ with $\tau \left( X
  \right) \leq \tau$, we have
  \begin{eqnarray}
    \left\| \Sigma_1^{1 / 2} X \Sigma_2^{1 / 2} \right\|_F & = & \left\|
    \left( \Sigma_2^{1 / 2} \otimes \Sigma_1^{1 / 2} \right) \tmop{vec} \left(
    X \right) \right\|_2\nn\\
    & \geq& \rho_{\tau} \left( \Sigma_2^{1 / 2} \otimes
    \Sigma_1^{1 / 2} \right) \left\| X \right\|_F .
  \end{eqnarray}
  As a consequence of Proposition \ref{pro:gaussian:estimate}, we obtain
  \begin{eqnarray}
&&    \hspace{-1cm}\left\| \mathcal{A}_{\Sigma_1, \Sigma_2} \left( X \right) \right\|_2 \nn\\
    &&
    \hspace{-1cm}\geq  \frac{1}{4} \rho_{\tau} \left( \Sigma_2^{1 / 2} \otimes \Sigma_1^{1
    / 2} \right) \left\| X \right\|_F\nn\\
    && \ \ - 3 \frac{\sqrt{\tmop{trace} \left(
    \Sigma_1 \right)} + \sqrt{\tmop{trace} \left( \Sigma_2 \right)}}{\sqrt{m}}
    \left\| X \right\|_{\ast}\nn\\
    &&    \hspace{-1cm} \geq \frac{1}{4} \rho_{\tau} \left( \Sigma_2^{1 / 2} \otimes
    \Sigma_1^{1 / 2} \right) \left\| X \right\|_F\nn\\
    &&    \hspace{-.5cm} - 3 \frac{\sqrt{\tmop{trace}
    \left( \Sigma_1 \right)} + \sqrt{\tmop{trace} \left( \Sigma_2
    \right)}}{\sqrt{m}} \sqrt{\tau} \left\| X \right\|_F .
  \end{eqnarray}
  The sample size condition
  \begin{eqnarray}
    m & \geq & c \frac{\tmop{trace} \left( \Sigma_1 \right) + \tmop{trace}
    \left( \Sigma_2 \right)}{\rho_{\tau}^2 \left( \Sigma_2^{1 / 2} \otimes
    \Sigma_1^{1 / 2} \right)} \tau
  \end{eqnarray}
   with $c = 2 \times 24^2$ then leads to
  \begin{eqnarray}
    \left\| \mathcal{A}_{\Sigma_1, \Sigma_2} \left( X \right) \right\|_2 
    \geq  \frac{1}{8} \rho_{\tau} \left( \Sigma_2^{1 / 2} \otimes \Sigma_1^{1
    / 2} \right) \left\| X \right\|_F
  \end{eqnarray}
for all $X$ such that $\tau \left( X \right) \leq \tau$,
  yielding the desired result.
\end{proof}

The proof of Proposition \ref{pro:gaussian:estimate} uses similar techniques
developed in \cite{Raskutti2010correlated} for correlated Gaussian design in the compressive sensing
setting. More specifically, we first show that (\ref{eqn:gaussian:estimate})
is true on the set $V \left( r \right) = \left\{ X \in \mathbbm{R}^{n_1 \times
n_2} : \left\| \Sigma_1^{1 / 2} X \Sigma_2^{1 / 2} \right\|_F = 1, \left\| X
\right\|_{\ast} \leq r \right\}$ for fixed $r > 0$ with high probability. A
peeling argument is then used to extend the result to all $X \in
\mathbbm{R}^{n_1 \times n_2}$. Refer to Appendix \ref{apx:gaussian:estimate} for more details.

\section{Fixed Point Characterization}\label{sec:fixedpoint}
In this section, we derive a fixed point characterization of $\rho_\tau(\A)$. Recall that the optimization problem defining $\rho_\tau$ is as follows:
\begin{eqnarray}\label{eqn:max_inf_Q_diamond}
\rho_\tau(\A) = \min_{Z} \frac{\|\A(Z)\|_2}{\|Z\|_\rmF} \text{\ s.t. \ } \frac{\|Z\|_*}{\|Z\|_\rmF} \leq \sqrt{\tau},
\end{eqnarray}
or equivalently,
\begin{eqnarray}\label{eqn:optimizerho}
  \frac{1}{\rho_\tau(\A)} = \max_{Z}\{ \|Z\|_\rmF: \|\A(Z)\|_2 \leq 1, \frac{\|Z\|_*}{\|Z\|_\rmF} \leq \sqrt{\tau}\}.
\end{eqnarray}

Denote by $\tau^* = \min_{Z: \A(Z) = 0} \tau(Z)$. 
For any $\tau \in (1, \tau^*)$, we define functions over $[0, \infty)$ parameterized by $\tau$:
\begin{eqnarray}
f_\tau(\eta; Y) = \max_Z\left\{\left<Y, Z\right>: \|\A(Z)\|_2 \leq 1, {\|Z\|_*} \leq \sqrt{\tau} \eta\right\},
\end{eqnarray}
for $Y \in \mathbb{S}^{n_1n_2-1}$ and 
\begin{eqnarray}
f_\tau(\eta) &=& \max_{Z} \left\{\|Z\|_\rmF: \|\A(Z)\|_2 \leq 1, {\|Z\|_*} \leq \sqrt{\tau} \eta\right\}\nn\\
 &=& \max_{Y,Z} \{\left<Y, Z\right>: \|\A(Z)\|_2 \leq 1, {\|Z\|_*} \leq \sqrt{\tau} \eta,\nn\\
 && \ \ \ \ \ \ \ \|Y\|_\rmF \leq 1\}\nn\\ 
  & = & \sup_{Y \in \mathbb{S}^{n_1n_2-1}} f_\tau(\eta; Y). \label{def:fs}
\end{eqnarray}
Here $\mathbb{S}^{n_1n_2-1}$ is the unit sphere in $(\R^{n_1\times n_2}, \|\cdot\|_\rmF)$. The continuity of $f_\tau(\eta; Y)$ with respect to $Y$, as established in Theorem \ref{thm:fix_fs}, implies that the supremum in \eqref{def:fs} can be replaced by maximum.
In the definition of $f_\tau(\eta)$, we basically replaced the $\|Z\|_\rmF$ in the denominator of the fractional constraint in \eqref{eqn:optimizerho} with $\eta$. 

For $\eta > 0$, it is easy to show that strong duality holds for the optimization problem defining $f_\tau(\eta; Y)$. As a consequence, we have the dual form of $f_\tau(\eta; Y)$:
\begin{eqnarray}\label{eqn:fsi_dual}
  f_\tau(\eta; Y) = \min_{\bs \lambda} \sqrt{\tau}\eta \|Y - \A^*(\bs \lambda)\| + \|\bs \lambda\|_2.
\end{eqnarray}

It turns out that the unique positive fixed point of $f_\tau(\eta)$ is exactly $1/\rho_\tau(\A)$, as shown by the following theorem. See Appendix \ref{app:pf:fix_fs} for the proof.

\begin{thm}\label{thm:fix_fs}
The functions $f_\tau(\eta; Y)$ and $f_\tau(\eta)$ have the following properties:
\begin{enumerate}
  \item $f_\tau(\eta; Y)$ and $f_\tau(\eta)$ are jointly continuous in $\tau, \eta$, and $Y$.
  \item $f_\tau(\eta; Y)$ and $f_\tau(\eta)$ are strictly increasing in $\eta$.
  \item $f_\tau(\eta; Y)$ is concave for each $Y \in \mathbb{S}^{n_1n_2-1}$.
\item $f_\tau(0) = 0$, $f_\tau(\eta) \geq s\eta > \eta$ for sufficiently small $\eta > 0$, and there exists $\rho < 1$ such that $f_\tau(\eta) < \rho\eta$ for sufficiently large $\eta$; the same holds for $f_\tau(\eta; Y)$ if $Y = \u\v^T$ with $\|\u\|_2 = 1$ and $\|\v\|_2 = 1$, and the existence of $\rho$ holds for all $f_\tau(\eta; Y)$.
    \item $f_\tau(\eta; Y)$ has unique positive fixed point for $Y = \u\v^T$ with $\|\u\|_2 = 1$ and $\|\v\|_2 = 1$; $f_\tau(\eta)$ has at least one positive fixed points 
  \begin{eqnarray}
  \eta^* = \max \bigcup_Y \{\eta: \eta = f_\tau(\eta; Y)\}.
  \end{eqnarray}  
  \item The positive fixed point $\eta^*$ of $f_\tau(\eta)$ is unique and satisfies
\begin{eqnarray}
\eta^* = \frac{1}{\rho_\tau(\A)}.
\end{eqnarray}  
    \item For $\eta \in (0, \eta^*)$, we have $f_\tau(\eta) > \eta$; and for $\eta \in (\eta^*, \infty)$, we have $f_\tau(\eta) < \eta$; the same statement holds also for $f_\tau(\eta; Y)$ if $Y = \u\v^T$ with $\|\u\|_2 = 1$ and $\|\v\|_2 = 1$.
  \item For any $\epsilon > 0$, there exists $\rho_1(\epsilon) > 1$ such that $f_\tau(\eta) > \rho_1(\epsilon) \eta$ as long as $0 < \eta \leq (1-\epsilon) \eta^*$; and there exists $\rho_2(\epsilon) < 1$ such that $f_\tau(\eta) < \rho_2(\epsilon) \eta$ as long as $\eta > (1+\epsilon) \eta^*$.
\end{enumerate}
\end{thm}

For sparsity recovery and block-sparsity recovery, the fixed point characterizations yield efficient algorithms to compute certain incoherence measures \cite{tang2011linf, tang2011blocksparsity}. We develop the fixed point characterization in this paper in the hope that it might also lead to a way to compute $\rho_\tau(\A)$. However, at this point, it is not clear how to compute or approximation $f_\tau(\eta)$ efficiently at a particular $\eta$.

\section{Conclusions}\label{sec:conclusions}

In this paper, the $\ell_*$-constrained minimal singular value of a measurement operator, which measures the invertibility of the measurement operator restricted to matrices with small $\ell_*$-rank, is proposed to quantify the stability of low-rank matrix reconstruction. The reconstruction errors of the matrix Basis Pursuit, the matrix Dantzig selector, and the matrix LASSO estimator are concisely bounded using the $\ell_*$-CMSV. We demonstrate that the $\ell_*$-CMSV is bounded away from zero with high probability for isotropic and subgaussian measurement operators, as long as the number of measurements is relatively large. We also show that for correlated Gaussian operator, the $\ell_*$-CMSV is lower bounded by that of its covariance matrix. Finally, we derive a fixed point characterization that is potentially useful for computing $\ell_*$-CMSV.

In the future work, we will design algorithms to efficiently compute or approximate the $\ell_*$-CMSV using the fixed point characterization. We also plan to extend the result for correlated Gaussian operator to subgaussian operators with correlation structure.

\appendices
\section{Proof of Proposition \ref{pro:errorcharacteristics}}\label{app:pf:pro:errorcharacteristics}
\label{app:pf:errorcharacteristics}
We need two lemmas about the properties of nuclear norms derived in \cite{Recht2010Nuclear}:

\begin{lm}\label{lm:nuclearadd}\hspace{-0.1cm}\emph{\cite[Lemma 2.3]{Recht2010Nuclear}}
Let $A$ and $B$ be matrices of the same dimensions. If $AB^T = 0$ and $A^TB = 0$ then $\|A + B\|_* = \|A\|_* + \|B\|_*$.
\end{lm}

\begin{lm}\label{lm:mDec}\hspace{-0.1cm}\emph{\cite[Lemma 3.4]{Recht2010Nuclear}}
Let $A$ and $B$ be matrices of the same dimensions. Then there exist matrices $B_1$ and $B_2$ such that
\begin{enumerate}
  \item $B = B_1 + B_2$
  \item $\rank(B_1) \leq 2\rank(A)$
  \item $AB_2^T = 0$ and $A^TB_2 = 0$
  \item $\left<B_1, B_2\right> = 0$.
\end{enumerate}
\end{lm}

\begin{proof}[Proof of Proposition \ref{pro:errorcharacteristics}] 
We first deal with the mBP and the mDS. We decompose the error matrix $B = H$ according to Lemma \ref{lm:mDec} with $A = X$, more explicitly, we have:
 \begin{enumerate}
  \item $H = H_0 + H_c$
  \item $\rank(H_0) \leq 2 \rank(X) = 2r$
  \item $XH_c^T = 0$ and $X^TH_c = 0$
  \item $\left<H_0, H_c\right> = 0$.
 \end{enumerate}
As observed by Recht \emph{et.al} in \cite{Recht2010Nuclear} (See also \cite{Candes2008RIP}, \cite{candes2009lowrank} and \cite{tang2011cmsv}),  the fact that $\|\hX\|_* = \|X + H\|_*$ is the minimum among all $Z$s satisfying the constraint in \eqref{mbp} implies that $\|H_c\|_*$ cannot be very large. To see this, we observe that
\begin{eqnarray}\label{x_min}
  \|X\|_* &\geq& \|X + H\|_*\nonumber\\
  & = & \|X + H_c + H_0\|_* \nonumber\\
  &\geq& \|X + H_c \|_* - \|H_0\|_*\nonumber\\
  & = & \|X\|_* + \|H_c\|_* - \|H_0\|_*.
\end{eqnarray}
Here, for the last equality we used Lemma \ref{lm:nuclearadd} and $XH_c^T = 0, X^TH_c = 0$. Therefore, we obtain
\begin{eqnarray}\label{cless0}
\|H_c\|_* \leq \|H_0\|_*,
\end{eqnarray}
which leads to
\begin{eqnarray}\label{h1h2}
\|H\|_*  &\leq& \|H_0\|_* + \|H_c\|_*\nonumber\\
 &\leq&  2\|H_0\|_* \nonumber\\
 &\leq& 2\sqrt{\rank(H_0)}\|H_0\|_\rmF\nn\\
 & = & 2\sqrt{2r} \|H\|_\rmF,
\end{eqnarray}
where for the next to the last inequality we used the fact that $\|H\|_* \leq \sqrt{\rank(H)}\|H\|_\rmF$, and for the last inequality we used the Pythagorean theorem $\|H\|_\rmF^2 = \|H_0\|_\rmF^2 + \|H_c\|_\rmF^2 \geq \|H_0\|_\rmF^2$ because $\left<H_0, H_c\right> = 0$. Inequality \eqref{h1h2} is equivalent to
\begin{eqnarray}\label{s_1}
\tau(H) \leq 8\ \rank(X) = 8r.
\end{eqnarray}

We now turn to the LASSO estimator \eqref{mlasso}. Suppose the noise $\w$ satisfies $\|\A^*(\w)\| \leq \kappa\mu$ for some small $\kappa > 0$. Because $\hX$ is a solution to \eqref{mlasso}, we have
\begin{eqnarray}
&&\frac{1}{2}\|\A(\hX) - \y\|_2^2 + \mu \|\hX\|_*\nn\\ &\leq& \frac{1}{2} \|\A(X) - \y\|_2^2 + \mu\|X\|_*.
\end{eqnarray}
Consequently, substituting $\y = \A(X) + \w$ yields
\begin{eqnarray}
\mu \|\hX\|_*
&\leq& \left<\A(\hX-X), \w\right> + \mu\|X\|_*\nn\\
& = & \left<\hX - X, \A^*(\w)\right> + \mu\|X\|_*.
\end{eqnarray}
Using the Cauchy-Swcharz type inequality, we get
\begin{eqnarray}
\mu \|\hX\|_* &\leq & \|\hX-X\|_*\|\A^*(\w)\| + \mu\|X\|_*\nn\\
& = & \kappa \mu \|H\|_* + \mu\|X\|_*,
\end{eqnarray}
which leads to
\begin{eqnarray}
  \|\hX\|_* &\leq & \kappa\|H\|_* + \|X\|_*.
\end{eqnarray}
Therefore, similar to the argument in \eqref{x_min} we have
\begin{eqnarray}
 \|X\|_* \geq  \|X\|_* + (1-\kappa)\|H_c\|_* - (1+\kappa)\|H_0\|_*
\end{eqnarray}
Consequently, we have
\begin{eqnarray}
  \|H_c\|_* &\leq& \frac{1+\kappa}{1-\kappa}\|H_0\|_*,
\end{eqnarray}
an inequality slightly worse than \eqref{cless0} for small $\kappa$.
Therefore, an argument similar to the one leading to \eqref{h1h2} yields
\begin{eqnarray}\label{lassoh1h2}
\|H\|_* \leq \frac{2}{1-\kappa} \sqrt{2r} \|H\|_\rmF,
\end{eqnarray}
or equivalently,
\begin{eqnarray}
  \tau(H) &\leq& \frac{8r}{(1-\kappa)^2}.
\end{eqnarray}
\end{proof}

\section{Proof of Theorem \ref{thm:errorbound}}\label{app:pf:thm:errorbound}

\begin{proof}
To prove Theorem \ref{thm:errorbound}, we only need to obtain upper bounds on $\|\A(H)\|_2$ and then invoke the definition of the $\ell_*$-CMSV. For mBP \eqref{mbp}, this is trivial as both $X$ and $\hX$ satisfy constraint $\|\y-\A(Z)\| \leq \varepsilon$ in \eqref{mbp}. Therefore, the triangle inequality yields
\begin{eqnarray}\label{Ah}
\|\A(H)\|_2 &=& \|\A(\hX - X)\|_2 \nonumber\\
&\leq& \|\A(\hX) - \y\|_2 + \|\y - \A(X)\|_2 \nonumber\\
&\leq& 2 \varepsilon.
\end{eqnarray}
It then follows from Definition \ref{def:lstarcmsv} that
\begin{eqnarray}
  \rho_{8r} \|H\|_\rmF  &\leq & \|\A(H)\|_2 \leq 2\varepsilon.
\end{eqnarray}
Hence, we get
\begin{eqnarray}
  \|\hX - X\|_\rmF &\leq& \frac{2\varepsilon}{\rho_{8r}}.
\end{eqnarray}
\\

For the mDS \eqref{mds}, the condition $\|\A^*(\w)\| \leq \mu$ and the constraint in \eqref{mds} yield
\begin{eqnarray}
  \|\A^*(\A(H))\| &\leq& 2\mu
\end{eqnarray}
because
\begin{eqnarray}
 \A^*(\w - \hat{\r}) &=& \A^*\left((\y-\A(X))-(\y-\A(\hX))\right) \nonumber\\
  &=& \A^*\left(\A(\hX) - \A(X)\right) \nn\\
  &=& \A^*(\A(H)),
\end{eqnarray}
where $\hat{\r} = \y - \A(\hX)$ is the residual corresponding to the mDS solution $\hX$.
Therefore, we obtain an upper bound on $\|\A(H)\|_2^2$ as follows:
\begin{eqnarray}\label{ds_Ah_ubd}
  \left<\A(H), \A(H)\right> &=& \left<H, \A^*(\A(H))\right>\nonumber\\
  & \leq &  \|H\|_*\|\A^*(\A(H))\|\nonumber\\
  & \leq & 2\mu \|H\|_*.
\end{eqnarray}
Equation \eqref{ds_Ah_ubd}, the definition of $\rho_{8r}$, and $\tau(H) \leq 8r$ together yield
\begin{eqnarray}
  \rho_{8r}^2 \|H\|_\rmF^2 &\leq& \left<\A(H), \A(H)\right> \nonumber\\
  &\leq& 2\mu \|H\|_* \nonumber\\
  &\leq& 2\mu \sqrt{8r} \|H\|_\rmF.
\end{eqnarray}
We conclude that
\begin{eqnarray}
  \|H\|_\rmF &\leq& \frac{4\sqrt{2r}}{\rho_{8r}^2} \mu.
\end{eqnarray}

Now we establish an upper bound on $\|\A(H)\|_2^2$ for the mLASSO \eqref{mlasso} using a  procedure similar to the one used for the mDS given above. First note that 
\begin{eqnarray}\label{lassoastara}
 &&\|\A^*(\A(H))\|\nn\\ &\leq& \|\A^*(\y-\A(X))\| + \|\A^*(\y - \A(\hX))\|\nn\\
 &\leq & \|\A^*(\w)\| + \|\A^*(\y - \A(\hX))\|\nn\\
 & = & \kappa \mu + \|\A^*(\y - \A(\hX))\|.
\end{eqnarray}

We follow the procedure in \cite{candes2009lowrank} (see also \cite{bickel2009simultaneous}) to estimate $\|\A^*(\y - \A(\hX))\|$. Since $\hX$ is the solution to \eqref{mlasso}, the optimality condition yields that
\begin{eqnarray}
\A^*(\y - \A(\hX)) \in \mu\partial \|\hX\|_*,
\end{eqnarray}
where $\partial \|\hX\|_*$ is the family of subgradient of $\|\cdot\|_*$ evaluated at $\hX$. According to \cite{recht2009exact}, if the singular value decomposition of $\hX$ is $U\Sigma V^T$, then we have
\begin{eqnarray}
&&\hspace{-1.5cm}\partial \|\hX\|_* = \{UV^T + W: \|W\| \leq 1,\nn\\
&&  \hspace{3cm} U^TW = 0, WV = 0\}.
\end{eqnarray}

As a consequence, we obtain $\A^*(\y - \A(\hX)) = \mu(UV^T + W)$ and
\begin{eqnarray}\label{lassoastarerro}
\|\A^*(\y - \A(\hX))\| &\leq& \|\mu(UV^T + W)\|\nn\\
& = & \mu.
\end{eqnarray}
We used $\|UV^T + W\| = 1$ because
\begin{eqnarray}
  &&\max_{\x: \|\x\|_2 = 1}\|(UV^T + W)\x\|\nn\\
  &=& \max_{\y:\|\y\|_2 = 1} \|(UV^T + W)V\y\| \leq 1.
\end{eqnarray}
Following the same lines in \eqref{ds_Ah_ubd}, we get
\begin{eqnarray}
\|\A(H)\|_2^2 \leq (\kappa + 1)\mu\|H\|_*.
\end{eqnarray}
Then, Equation \eqref{lassoh1h2}, \eqref{lassoastara} and \eqref{lassoastarerro}
\begin{eqnarray}
  &&\rho_{\frac{8r}{(1-\kappa)^2}}^2\|H\|_\rmF^2 \leq  \|\A(H)\|_2^2\nn\\
& \leq & (\kappa + 1)\mu \frac{\sqrt{8r}}{1-\kappa} \|H\|_\rmF.
\end{eqnarray}
As a consequence, the error bound \eqref{bd:mlasso} holds.
\end{proof}

\section{Proof of Theorem \ref{thm:fix_fs}}\label{app:pf:fix_fs}
\begin{proof}
\begin{enumerate}
\item Since in the optimization problem defining $f_\tau(\eta; Y)$, the objective function $\left<Y, Z\right>$ is jointly continous in $\tau, \eta$ and $Y$, and the constraint correspondence
\begin{eqnarray}
 && \hspace{-1.2cm}C(\tau, \eta): [0, \infty) \twoheadrightarrow \R^{n_1\times n_2}\nonumber\\
  && \hspace{-1.2cm}\eta \mapsto \left\{Z: \|\A(Z)\|_2 \leq 1, {\|Z\|_*} \leq \sqrt{\tau} \eta\right\}
\end{eqnarray}
is compact-valued and continuous (both upper and lower hemicontinuous), according to Berge's Maximum Theorem \cite{Berge1997maximum}, the optimal value function $f_\tau(\eta; Y)$ is jointly continuous in $\tau, \eta$ and $Y$. The continuity of $f_\tau(\eta)$ can be proved in a similar manner.

\item  

To show the strict increasing property, suppose $0 < \eta_1 < \eta_2$ and the dual variable ${\bs \lambda}_2^*$ achieves $f_\tau(\eta_2; Y)$ in \eqref{eqn:fsi_dual}. Then we have
\begin{eqnarray}
  f_\tau(\eta_1; Y) &\leq&  \sqrt{\tau}\eta _1 \|Y - \A({\bs \lambda}_2^*)\| + \|{\bs \lambda}_2^*\|_2\nonumber\\
  &<& \sqrt{\tau}\eta _2 \|Y - \A({\bs \lambda}_2^*)\| + \|{\bs \lambda}_2^*\|_2\nonumber\\
  &=& f_\tau(\eta_2; Y).
\end{eqnarray}
The case for $\eta_1 = 0$ is proved by continuity, and the strict increasing of $f_\tau(\eta)$ follows immediately.

\item The concavity of $f_\tau(\eta; Y)$ follows from the dual representation \eqref{eqn:fsi_dual} and the fact that $f_\tau(\eta; Y)$ is the minimization of a function of variables $\eta$ and $\bs \lambda$, and when $\bs\lambda$, the variable to be minimized, is fixed, the function is linear in $\eta$.

\item Next we show that when $\eta > 0$ is sufficiently small $f_\tau(\eta; Y) \geq \sqrt{\tau} \eta$ if $Y = \u\v^T$ with $\|\u\|_2 = 1$ and $\|\v\|_2 = 1$. Taking $Z = \sqrt{\tau} \eta \u\v^T$, we have $\|Z\|_* = \sqrt{\tau}\eta$ and $\left<Y, Z\right> = \sqrt{\tau}\eta > \eta$ (recall $\tau \in (1, \tau^*)$). In addition, when $0 < \eta \leq 1/(\sqrt{\tau}\|\A(\u\v^T)\|_2)$, we also have $\|\A(Z)\|_2 \leq 1$. Therefore, for sufficiently small $\eta$, we have $f_\tau(\eta; Y) \geq \sqrt{\tau} \eta > \eta$. Clearly, $f_\tau(\eta) = \max_Y f_\tau(\eta; Y) \geq \sqrt{\tau} \eta > \eta$ for such $\eta$.

Recall that
\begin{eqnarray}
\tau^* = \min_Z \frac{\|Z\|_*^2}{\|Z\|_\rmF^2} \text{\ subject to \ } \A(Z) = 0,
\end{eqnarray}
or equivalently $\frac{1}{\sqrt{\tau^*}}$ is
\begin{eqnarray}
\max_{Z, Y: \|Y\|_\rmF = 1}\{ \left<Y, Z\right>: \A(Z) = 0, \|Z\|_* \leq 1\}.
\end{eqnarray}
Since the dual program of 
\begin{eqnarray}
\max_Z \left<Y, Z\right> \text{\ s.t. \ } \A(Z) = 0, \|Z\|_* \leq 1
\end{eqnarray}
is 
\begin{eqnarray}
\max_{\bs \lambda} \|Y - \A^*(\bs \lambda)\|, 
\end{eqnarray}
we have
\begin{eqnarray}
  \frac{1}{\sqrt{\tau^*}} &=&  \max_{Y: \|Y\|_\rmF = 1} \min_{{\bs \lambda}} \|Y - \A^*({\bs \lambda})\|.
\end{eqnarray}
Suppose ${\bs \lambda}_Y^*$ is the optimal solution for each $\min_{{\bs \lambda}}\|Y - \A^*({\bs \lambda})\|$. For each $Y$, we then have
\begin{eqnarray}\label{eqn:taustarbd}
  \frac{1}{\sqrt{\tau^*}} &\geq & \|Y - \A^*({\bs \lambda}_Y^*)\|,
\end{eqnarray}
which implies
\begin{eqnarray}
  f_\tau(\eta; Y) &=& \min_{{\bs \lambda}} \sqrt{\tau}\eta  \|Y - \A^*({\bs \lambda})\| + \|{\bs \lambda}\|_2 \nonumber\\
  &\leq & \sqrt{\tau}\eta  \|Y - \A^*({\bs \lambda}_Y^*)\| + \|{\bs \lambda}_Y^*\|_2 \nonumber\\
  &\leq & \sqrt{\frac{\tau}{\tau^*}} \eta + \|{\bs \lambda}_Y^*\|_2.
\end{eqnarray}
As a consequence, we obtain
\begin{eqnarray}
f_\tau(\eta) &=& \max_Y f_\tau(\eta; Y)\nn\\
& \leq &\sqrt{\frac{\tau}{\tau_*}} \eta + \sup_Y \|{\bs \lambda}_Y^*\|_2.
\end{eqnarray}

Viewing $\A^*: (\R^m, \|\cdot\|_2) \rightarrow (\R^{n_1\times n_2}, \|\cdot\|)$ as an operator, 
we obtain that $\theta_\A \df \inf_{\bs \lambda \neq 0} \|\A^*(\bs \lambda)\|/\|\bs \lambda\|_2 > 0$ when $\{A^k\}_{k=1}^m$ are linearly independent, because $\A^*(\bs \lambda) \neq 0$ for $\bs \lambda \neq 0$. Triangle inequality and \eqref{eqn:taustarbd} imply
\begin{eqnarray}
\frac{1}{\sqrt{\tau^*}} + 1 &\geq& \frac{1}{\sqrt{\tau^*}} + \|Y\| \nn\\
&\geq& \|\A^*(\bs \lambda_Y)\| \geq \theta_\A \|\bs \lambda_Y\|_2.
\end{eqnarray}
Therefore, the quantify $\sup \|{\bs \lambda}_Y^*\|_2$ is finite. 
Pick $\rho \in (\sqrt{\tau/\tau_*}, 1)$. Then, we have the following when $\eta > \sup_Y \|{\bs \lambda}_Y^*\|_2/(\rho - \sqrt{\tau/\tau_*})$:
\begin{eqnarray}
  f_\tau(\eta; Y) &\leq& \rho \eta, \forall Y \in \mathbb{S}^{n_1n_2 - 1} \text{\ and \ } \nn\\
  f_\tau(\eta) &\leq& \rho \eta.
\end{eqnarray}

\item 
Properties 1) and 4) imply that $f_\tau(\eta; Y)$ has at least one positive fixed point for $Y = \u\v^T$. (Interestingly, 2) and 4) also imply the existence of a positive fixed point, see \cite{tarski1955fixedpoint}.) The positive fixed point for such $f_\tau(\eta; Y)$ is also unique. Suppose there are two fixed points $0 < \eta_1^* < \eta_2^*$. Pick $\eta_0$ small enough such that $f_\tau(\eta_0; Y) > \eta_0 > 0$ and $\eta_0 < \eta_1^*$. Then $\eta_1^* = \lambda \eta_0 + (1-\lambda)\eta_2^*$ for some $\lambda \in (0, 1)$, which implies that $f_\tau(\eta_1^*; Y) \geq \lambda f_\tau(\eta_0; Y) + (1-\lambda) f_\tau(\eta_2^*; Y) > \lambda \eta_0 + (1-\lambda) \eta_2^* = \eta_1^*$ due to the concavity, contradicting $\eta_1^* = f_\tau(\eta_1^*; Y)$.

 The set of positive fixed point for $f_\tau(\eta)$, $\{\eta \in (0, \infty): \eta = f_\tau(\eta) = \max_Y f_\tau(\eta; Y)\}$, is a subset of $\{\eta \in (0, \infty):  \eta = f_\tau(\eta; Y) \text{\ for some \ } Y \}$, which is non-empty due to the existence of fixed points for $f_\tau(\eta; Y)$ with $Y = \u\v^T$. We argue that $\eta^*$ is the supremum of 
    \begin{eqnarray}
  \{\eta \in (0, \infty):  \eta = f_\tau(\eta; Y) \text{\ for some \ } Y \}
    \end{eqnarray}
is the unique positive fixed point for $f_\tau(\eta)$.

First of all, $\eta^*$ must be finite as all the fixed points for $f_\tau(\eta; Y)$ are less than $\sup_Y \|{\bs \lambda}_Y^*\|_2/(\rho - \sqrt{\tau/\tau_*})$ according to the proof of property 4). Second, $\eta^*$ is a fixed point for some $f_\tau(\eta; Y^*)$, namely, the supremum is achievable and can be replaced by maximum. To see this, we construct $\{\eta_k\}_{k=1}^\infty$ converging to $\eta^*$ using the the definition of $\eta^*$, and corresponing $\{Y_k\}_{k=1}^\infty$ (\emph{i.e.} $\eta_k$ is a fixed point of $f_\tau(\eta; Y_k)$) converging to some $Y^*$ using the compactness of $\mathbb{S}^{n_1n_2-1}$.  The joint continuity of $f_\tau(\eta; Y)$ in both $\eta$ and $Y$ implies
\begin{eqnarray}
\eta^* &&= \lim_{k \rightarrow \infty} \eta_k = \lim_{k\rightarrow \infty} f_\tau(\eta_k; Y_k) \nn\\
&&= f_\tau(\eta^*; Y^*).
\end{eqnarray}

We proceed to show that $\eta^*$ is a fixed point of $f_\tau(\eta)$. It suffices to show that $\max_Y f_\tau(\eta^*; Y) = f_\tau(\eta^*; Y^*)$. If this is not the case, there exists $Y_1 \neq Y^*$ such that $f_\tau(\eta^*; Y_1) > f_\tau(\eta^*; Y^*) = \eta^*$. The continuity of $f_\tau(\eta; Y_1)$ and the property 4) imply that there exists $\eta > \eta^*$ with $f_\tau(\eta; Y_1) = \eta$, contradicting the definition of $\eta^*$.


\item Next we show $\eta^*  = \gamma^* \df 1/\rho_\tau(\A)$ for any positive fixed point of $f_\tau(\eta)$, hence the uniqueness. We first prove $\gamma^* \geq \eta^*$ for any fixed point $\eta^* = f_\tau(\eta^*)$. Suppose $Z^*$ achieves the optimization problem defining $f_\tau(\eta^*)$, then we have
\begin{eqnarray}
\eta^* = f_\tau(\eta^*) = \|Z^*\|_\rmF,  \|\A(Z^*)\|_2 \leq 1,\\ \text{and\ } \|Z^*\|_* \leq \sqrt{\tau}\eta^*.
\end{eqnarray}
Since $\|Z^*\|_*/\|Z^*\|_\rmF \leq \sqrt{\tau}\eta^*/\eta^* \leq \sqrt{\tau}$, we have
\begin{eqnarray}
  \gamma^* &\geq& \frac{\|Z^*\|_\rmF}{\|\A(Z^*)\|_2} \geq \eta^*.
\end{eqnarray}

If $\eta^* < \gamma^*$, we define $\eta_0 = (\eta^* + \gamma^*)/2$ and
\begin{eqnarray}
\hskip -1cm &&Z^{\mathrm{c}} = \mathrm{argmax}_{Z}{\frac{\sqrt{\tau}\|Z\|_\rmF}{\|Z\|_*}}\nn\\
&& \text{\ s.t. \ } \|\A(Z)\|_2 \leq 1, \|Z\|_\rmF \geq \eta_0,\label{eqn:defzc}\\
\hskip -1cm  &&\rho = {\frac{\sqrt{\tau}\|Z^{\mathrm{c}}\|_\rmF}{\|Z^{\mathrm{c}}\|_*}}.\label{eqn:defrho}
\end{eqnarray}
Suppose $Z^{**}$ with $\|\A(Z^{**})\|_2 = 1$ achieves the optimum of the optimization \eqref{eqn:max_inf_Q_diamond} defining $\gamma^* = 1/\rho_\tau(\A)$. Clearly, $\|Z^{**}\|_\rmF = \gamma^* > \eta_0$, which implies $Z^{**}$ is a feasible point of the optimization problem \eqref{eqn:defzc}  defining $Z^{\mathrm{c}}$ and $\rho$. As a consequence, we have
\begin{eqnarray}
  \rho \geq {\frac{\sqrt{\tau}\|Z^{**}\|_\rmF}{\|Z^{**}\|_*}} \geq 1.
\end{eqnarray}

\begin{figure}[h!t]
\hskip -0cm
\centering
\includegraphics[width = 0.3\textwidth, trim = 0mm 0mm 0mm 0mm, clip]{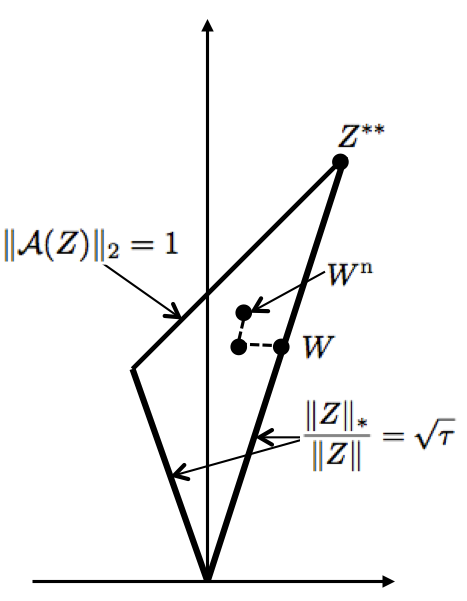}
\caption{Illustration of the proof for $\rho > 1$.}
\label{fig:proof}
\end{figure}%

Actually we will show that $\rho > 1$. If $\|Z^{**}\|_* < \sqrt{\tau}\|Z^{**}\|_\rmF$, we are done. If not (\emph{i.e.}, $\|Z^{**}\|_* = \sqrt{\tau}\|Z^{**}\|_\rmF$), as illustrated in Figure \ref{fig:proof}, we consider $W = \frac{\eta_0}{\gamma^*}Z^{**}$, which satisfies
\begin{eqnarray}
  &&\|\A(W)\|_2 = \frac{\eta_0}{\gamma^*} < 1,\\
  &&\|W\|_\rmF= \eta_0, \text{\  and \ }\\
  &&\|W\|_* = \sqrt{\tau}\eta_0.
\end{eqnarray}

Suppose $\bs \sigma$ is the singular value vector of $W$. To get $W^{\mathrm{n}}$ as shown in Figure \ref{fig:proof}, pick the smallest non-zero singular value, and scale it by a small positive constant $\kappa$ less than $1$. Because $\tau > 1$, $\bs \sigma$ has more than one non-zero components, elementary mathematics then show that this first scaling will decrease the ratio $\|\bs \sigma\|_1/\|\bs \sigma\|_2$. We then scale the entire vector $\bs \sigma$ so that its $\ell_2$ norm restores to its original value. This latter process of course does not change the ratio $\|\bs \sigma\|_1/\|\bs \sigma\|_2$. 

If the scaling constant $\kappa$ is close enough to $1$, $\|\A(W^{\mathrm{n}})\|_2$ will remain less than 1 due to continuity. But the good news is that the ratio $\|{\bs \sigma}\|_{1}/\|\bs \sigma\|_2 = \|W^\mathrm{n}\|_*/\|W^\mathrm{n}\|_\rmF$ decreases, and hence $\rho \geq \frac{\sqrt{\tau}\|{W}^{\mathrm{n}}\|_\rmF}{\|{W}^{\mathrm{n}}\|_*}$ becomes greater than 1.

Now we proceed to obtain a contradiction that $f_\tau(\eta^*) > \eta^*$. If $\|Z^{\mathrm{c}}\|_* \leq \sqrt{\tau}\cdot \eta^*$, then it is a feasible point of
\begin{eqnarray}\label{eqn:subt0}
  &&\hspace{-1cm}\max_{Z} \|Z\|_\rmF\nn\\
  &&\hspace{-1cm}\text{\ s.t. \ } \|\A(Z)\|_2 \leq 1, \|Z\|_* \leq \sqrt{\tau}\cdot \eta^*.
\end{eqnarray}
As a consequence, $f_\tau(\eta^*) \geq \|Z^{\mathrm{c}}\|_\rmF\geq \eta_0 > \eta^*$, contradicting $\eta^*$ is a fixed point and we are done. If this is not the case, \emph{i.e.}, $\|Z^{\mathrm{c}}\|_* > \sqrt{\tau}\cdot \eta^*$, we define a new point
\begin{eqnarray}
  Z^{\mathrm{n}} = \tau Z^{\mathrm{c}}
\end{eqnarray}
with
\begin{eqnarray}
  \tau = \frac{\sqrt{\tau}\cdot \eta^*}{\|Z^\mathrm{c}\|_*} < 1.
\end{eqnarray}
Note that $Z^{\mathrm{n}}$ is a feasible point of the optimization problem defining $f_\tau(\eta^*)$ since
\begin{eqnarray}
&&\|\A(Z^{\mathrm{n}})\|_2 = \tau \|\A(Z^{\mathrm{c}})\|_2 < 1,\\
&&\|Z^{\mathrm{n}}\|_* = \tau \|Z^{\mathrm{c}}\|_* = \sqrt{\tau}\cdot \eta^*.
\end{eqnarray}
Furthermore, we have
\begin{eqnarray}
  \|Z^{\mathrm{n}}\|_\rmF= \tau \|Z^{\mathrm{c}}\|_\rmF= \rho \eta^*.
\end{eqnarray}
As a consequence, we obtain a contradiction
\begin{eqnarray}
  f_\tau(\eta^*) &\geq& \rho \eta^* > \eta^*.
\end{eqnarray}
\begin{figure}[h!t]
\hskip -0cm
\centering
\includegraphics[width = 0.3\textwidth, trim = 0mm 0mm 0mm 0mm, clip]{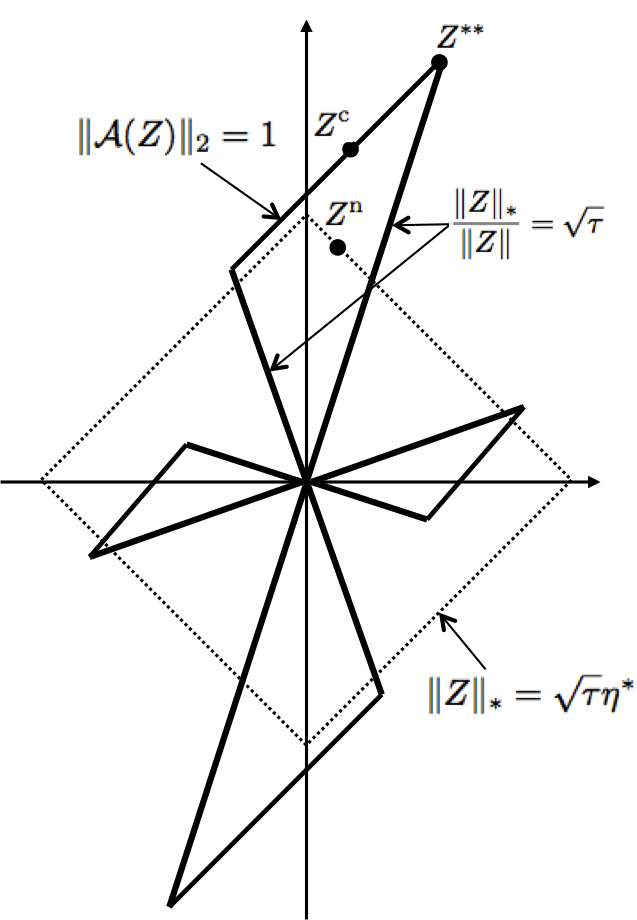}
\caption{Illustration of the proof for $f_\tau(\eta^*) \geq \rho\eta^*$.}
\label{fig:proof1}
\end{figure}%
Therefore, for the fixed point $\eta^*$, we have $\eta^* = \gamma^* = 1/\rho_\tau(\A)$.

\item This property simply follows from the continuity, the uniqueness, and property 4).
\item We use contradiction to show the existence of $\rho_1(\epsilon)$ in 8). In view of 4), we need only to show the existence of such a $\rho_1(\epsilon)$ that works for $\eta_L \leq \eta \leq (1-\epsilon)\eta^*$ where $\eta_L = \sup\{\eta: f_\tau(\xi) \geq \sqrt{\tau}\xi, \forall 0 < \xi \leq \eta\}$. Suppose otherwise, we then construct sequences $\{\eta^{(k)}\}_{k=1}^\infty \subset [\eta_L, (1-\epsilon)\eta^*]$ and $\{\rho_1^{(k)}\}_{k=1}^\infty \subset (1, \infty)$ with
\begin{eqnarray}
&& \lim_{k\rightarrow \infty} \rho_1^{(k)} = 1,\nonumber\\
&& f_\tau(\eta^{(k)}) \leq \rho^{(k)} \eta^{(k)}.
\end{eqnarray}
Due to the compactness of $[\eta_L, (1-\epsilon)\eta^*]$, there must exist a subsequence $\{\eta^{(k_l)}\}_{l=1}^\infty$ of $\{\eta^{(k)}\}$ such that $\lim_{l\rightarrow \infty}\eta^{(k_l)} = \eta_{\mathrm{lim}}$ for some $\eta_{\mathrm{lim}} \in [\eta_L, (1-\epsilon)\eta^*]$. As a consequence of the continuity of $f_\tau(\eta)$, we have

\begin{eqnarray}
f_\tau(\eta_{\mathrm{lim}}) &= & \lim_{l\rightarrow \infty} f_\tau(\eta^{(k_l)})\nn\\
& \leq & \lim_{l\rightarrow \infty} \rho_1^{(k_l)} \eta^{(k_l)} = \eta_{\mathrm{lim}}.
\end{eqnarray}
Again due to the continuity of $f_\tau(\eta)$ and the fact that $f_\tau(\eta) > \eta$ for $\eta < \eta_L$, there exists $\eta_c \in [\eta_L, \eta_{\mathrm{lim}}]$ such that
\begin{eqnarray}
  f_\tau(\eta_c) &=& \eta_c,
\end{eqnarray}
contradicting the uniqueness of the fixed point for $f_\tau(\eta)$. The existence of $\rho_2(\epsilon)$ can be proved in a similar manner.
\end{enumerate}
\end{proof}

\section{Proof of Proposition \ref{pro:gaussian:estimate}}\label{apx:gaussian:estimate}
In this appendix, we provide a proof of Proposition
\ref{pro:gaussian:estimate}.
For any $r$ such that the set $V \left( r \right) = \left\{ X \in
\mathbbm{R}^{n_1 \times n_2} : \left\| \Sigma_1^{1 / 2} X \Sigma_2^{1 / 2}
\right\|_F = 1, \left\| X \right\|_{\ast} \leq r \right\}$ is non-empty,
define a random variable
\begin{eqnarray}
  &&M \left( r, \mathcal{A}_{\Sigma_1, \Sigma_2} \right)  :=  1 - \inf_{X
  \in V \left( r \right)} \left\| \mathcal{A}_{\Sigma_1, \Sigma_2} \left( X
  \right) \right\|_2 \nn\\
  &=& \sup_{X \in V \left( r \right)} \left\{ 1 - \left\|
  \mathcal{A}_{\Sigma_1, \Sigma_2} \left( X \right) \right\|_2 \right\}
\end{eqnarray}
We first show that
\begin{eqnarray}
 &&\hspace{-1cm} \mathbbm{E}M \left( r, \mathcal{A}_{\Sigma_1, \Sigma_2} \right)\nn\\
  &&\hspace{-1cm}\leq
  \frac{1}{4} +  \frac{\sqrt{\tmop{trace} \left( \Sigma_1 \right)} +
  \sqrt{\tmop{trace} \left( \Sigma_2 \right)}}{\sqrt{m}} r.
\end{eqnarray}
To this end, we define two Gaussian processes indexed by $\tmmathbf{u} \in
  S^{m - 1}, X \in V \left( r \right)$
\begin{eqnarray}
Y_{\tmmathbf{u}, X} & =&  \left\langle \tmmathbf{u}, \mathcal{A}_{\Sigma_1,
  \Sigma_2} \left( X \right) \right\rangle \nn\\
  &=& \sum_{k, i, j} u_k G^k_{i j}
  \left( \Sigma_1^{1 / 2} X \Sigma_2^{1 / 2} \right)_{i j}, \\
  Z_{\tmmathbf{u}, X} & = & \left\langle \tmmathbf{u}, \tmmathbf{g}
  \right\rangle + \left\langle \Sigma_1^{1 / 2} X \Sigma_2^{1 / 2}, G
  \right\rangle,\end{eqnarray}
where $S^{m - 1}$ is the unit sphere in $\mathbbm{R}^m$, and $g_k, G_{i j}, G_{i j}^k$ are i.i.d. $\mathcal{N} \left( 0, 1 / m \right)$ random
variables. Observe that
\begin{eqnarray}
  &&\sup_{X \in V \left( r \right)} \inf_{\tmmathbf{u} \in S^{m - 1}}
  Y_{\tmmathbf{u}, X} = - \inf_{X \in V \left( r \right)} \sup_{\tmmathbf{u}
  \in S^{m - 1}} Y_{\tmmathbf{u}, X}\nn\\
   & = & - \inf_{X \in V \left( r \right)}
  \left\| \mathcal{A}_{\Sigma_1, \Sigma_2} \left( X \right) \right\|_2 .
\end{eqnarray}
We apply Gordon's comparision theorem \cite[Chapter 3.1]{ledoux1991probability} which states that if
\begin{eqnarray}
  \hspace{-1cm}\mathbbm{E} \left( Y_{\tmmathbf{u}, X} - Y_{\tmmathbf{u}', X'} \right)^2 
  \leq  \mathbbm{E} \left( Z_{\tmmathbf{u}, X} - Z_{\tmmathbf{u}', X'}
  \right)^2,   \label{eqn:gordon:cond}
\end{eqnarray}
for all $\left( \tmmathbf{u}, X \right)$, 
  $\left( \tmmathbf{u}', X' \right) \tmop{in} S^{m - 1} \times V \left( r
  \right)$ and the inequality becomes equality when $X = X'$, then
\begin{eqnarray}
  \hspace{-.2cm}\mathbbm{E} \sup_{X \in V \left( r \right)} \inf_{\tmmathbf{u} \in S^{m -
  1}} Y_{\tmmathbf{u}, X}  \leq  \mathbbm{E} \sup_{X \in V \left( r \right)}
  \inf_{\tmmathbf{u} \in S^{m - 1}} Z_{\tmmathbf{u}, X} .
\end{eqnarray}
Elementary calculations show that
\begin{eqnarray}
 && \mathbbm{E} \left( Y_{\tmmathbf{u}, X} - Y_{\tmmathbf{u}', X'} \right)^2\nn\\ & =
  & \frac{1}{m} \sum_{k, i, j} \left( u_k \left( \Sigma_1^{1 / 2} X
  \Sigma_2^{1 / 2} \right)_{i j} - u_k' \left( \Sigma_1^{1 / 2} X' \Sigma_2^{1
  / 2} \right)_{i j} \right)^2\nn\\
  & \leq & \frac{1}{m} \sum_k \left( u_k - u_k' \right)^2\nn\\
   &&+ \frac{1}{m}\sum_{i,
  j} \left( \left( \Sigma_1^{1 / 2} X \Sigma_2^{1 / 2} \right)_{i j} - \left(
  \Sigma_1^{1 / 2} X' \Sigma_2^{1 / 2} \right)_{i j} \right)^2 \nn\\
  & = & \mathbbm{E} \left( Z_{\tmmathbf{u}, X} - Z_{\tmmathbf{u}', X'}
  \right)^2
\end{eqnarray}
and
\begin{eqnarray}
  \mathbbm{E} \left( Y_{\tmmathbf{u}, X} - Y_{\tmmathbf{u}', X} \right)^2 & =
  & \mathbbm{E} \left( Z_{\tmmathbf{u}, X} - Z_{\tmmathbf{u}', X} \right)^2 \nn\\
  &=&
  \frac{1}{m} \sum_k \left( u_k - u_k' \right)^2 .
\end{eqnarray}
As a consequence, we upper bound the expectation of $M \left( r,
\mathcal{A}_{\Sigma_1, \Sigma_2} \right)$ as follows:
\begin{eqnarray}
  &&\hspace{-1cm}\mathbbm{E}M \left( r, \mathcal{A}_{\Sigma_1, \Sigma_2} \right)\nn\\ &&\hspace{-1cm} =   1
  +\mathbbm{E} \sup_{X \in V \left( r \right)} \inf_{\tmmathbf{u} \in S^{m -
  1}} Y_{\tmmathbf{u}, X} \nonumber\\
  &&\hspace{-1cm} \leq  1 +\mathbbm{E} \sup_{X \in V \left( r \right)} \inf_{\tmmathbf{u}
  \in S^{m - 1}} Z_{\tmmathbf{u}, X} \nonumber\\
  &&\hspace{-1cm}\leq 1 -\mathbbm{E} \left\| \tmmathbf{g} \right\|_2 +\mathbbm{E}
  \sup_{X \in V \left( r \right)} \left| \left\langle \Sigma_1^{1 / 2} X
  \Sigma_2^{1 / 2}, G \right\rangle \right| .  \label{eqn:EM}
\end{eqnarray}
For $\mathbbm{E} \left\| \tmmathbf{g} \right\|_2$, we use a simple lower bound
$\mathbbm{E} \left\| \tmmathbf{g} \right\|_2 \geq \frac{3}{4}
\sqrt{\frac{m}{m}} = \frac{3}{4}$ \cite{Raskutti2010correlated}. For the last term in (\ref{eqn:EM}), by
the definition of $V \left( r \right)$, we have
\begin{eqnarray}
  &&\mathbbm{E} \sup_{X \in V \left( r \right)} \left| \left\langle \Sigma_1^{1
  / 2} X \Sigma_2^{1 / 2}, G \right\rangle \right|\nn\\
  & = & \mathbbm{E} \sup_{X
  \in V \left( r \right)} \left| \left\langle X, \Sigma_1^{1 / 2} G
  \Sigma_2^{1 / 2} \right\rangle \right|\nn\\
  & \leq & \mathbbm{E} \left\| X \right\|_{\ast} \left\| \Sigma_1^{1 / 2} G
  \Sigma_2^{1 / 2} \right\|\nn\\
  & \leq & r\mathbbm{E} \left\| \Sigma_1^{1 / 2} G \Sigma_2^{1 / 2} \right\|
  .
\end{eqnarray}
The problem then boils down to estimating $\mathbbm{E} \left\| \Sigma_1^{1 /
2} G \Sigma_2^{1 / 2} \right\|$, which is achieved by applying Slepian's
comparison theorem \cite[Chapter 3.1]{ledoux1991probability}  to the following two Gaussian processes indexed by $\tmmathbf{u} \in S^{n_1 - 1}, \tmmathbf{v}
  \in S^{n_2 - 1}$:
\begin{eqnarray}
  Y_{\tmmathbf{u}, \tmmathbf{v}} & = & \tmmathbf{u}^T \Sigma_1^{1 / 2} G
  \Sigma_2^{1 / 2} \tmmathbf{v}, \\
  Z_{\tmmathbf{u}, \tmmathbf{v}} & = & \tmmathbf{u}^T \Sigma_1^{1 / 2}
  \tmmathbf{g}+\tmmathbf{v}^T \Sigma_2^{1 / 2} \tmmathbf{h}, 
\end{eqnarray}
where $\tmmathbf{g} \sim \mathcal{N} \left( 0, \frac{1}{m} I_{n_1} \right)$
and $\tmmathbf{h} \sim \mathcal{N} \left( 0, \frac{1}{m} I_{n_2} \right)$. It
is easy to verify that the variance condition for Slepian's comparison theorem
holds:
\begin{eqnarray}
\mathbbm{E} \left( Y_{\tmmathbf{u}, \tmmathbf{v}} - Y_{\tmmathbf{u}',
  \tmmathbf{v}'} \right)^2
 \leq\mathbbm{E} \left( Z_{\tmmathbf{u}, \tmmathbf{v}} - Z_{\tmmathbf{u}',
  \tmmathbf{v}'} \right)^2 .
\end{eqnarray}
Slepian's inequality $\mathbbm{E} \sup_{\tmmathbf{u}, \tmmathbf{v}}
Y_{\tmmathbf{u}, \tmmathbf{v}} \leq \mathbbm{E} \sup_{\tmmathbf{u},
\tmmathbf{v}} Z_{\tmmathbf{u}, \tmmathbf{v}}$ and Jensen's inequality then
imply
\begin{eqnarray}
  &&\mathbbm{E} \left\| \Sigma_1^{1 / 2} G \Sigma_2^{1 / 2} \right\| \nn\\
  & \leq &
  \mathbbm{E} \left\| \Sigma_1^{1 / 2} \tmmathbf{g} \right\|_2 +\mathbbm{E}
  \left\| \Sigma_2^{1 / 2} \tmmathbf{h} \right\|_2\nn\\
  & \leq & \sqrt{\mathbbm{E} \left\| \Sigma_1^{1 / 2} \tmmathbf{g}
  \right\|_2^2} + \sqrt{\mathbbm{E} \left\| \Sigma_2^{1 / 2} \tmmathbf{h}
  \right\|_2^2}\nn\\
  & = & \frac{1}{\sqrt{m}} \left( \sqrt{\tmop{trace} \left( \Sigma_1 \right)}
  + \sqrt{\tmop{trace} \left( \Sigma_2 \right)} \right) .
\end{eqnarray}
Plugging back into (\ref{eqn:EM}) yields
\begin{eqnarray}
  &&\hspace{-1cm}\mathbbm{E}M \left( r, \mathcal{A}_{\Sigma_1, \Sigma_2} \right)\nn\\ 
  &&\hspace{-1cm} \leq
  \frac{1}{4} + \frac{\sqrt{\tmop{trace} \left( \Sigma_1 \right)} +
  \sqrt{\tmop{trace} \left( \Sigma_2 \right)}}{\sqrt{m}} r.
\end{eqnarray}

We next obtain a high probability result using concentration of measure in
Gauss space for Lipschitz functions. Denote $\tmmathbf{e}_k$ as the $k$th
canonical basis vector. The following manipulation of
\begin{eqnarray}
&& h \left( G^1,
\cdots, G^m \right)\nn\\
&& : =  \sup_{X \in V \left( r \right)}
\left( 1 - \left\| \sum_k \left\langle G^k, \Sigma_1^{1 / 2} X \Sigma_2^{1 /
2} \right\rangle \tmmathbf{e}_k \right\|_2 \right) \nn\\
&&= M \left( r,
\mathcal{A}_{\Sigma_1, \Sigma_2} \right)
\end{eqnarray}
 as a function of $\left\{ G^1, \cdots, G^m \right\}$:
\begin{eqnarray}
  &&h \left( G^1, \cdots, G^m \right)\nn\\
  & \leq & 1 + \sup_{X \in V \left( r \right)} - \left\| \sum_k \left\langle
  G'^k, \Sigma_1^{1 / 2} X \Sigma_2^{1 / 2} \right\rangle \tmmathbf{e}_k
  \right\|_2 \nn\\
  && + \sup_{X \in V \left( r \right)} \left\| \sum_k \left\langle G^k
  - G'^k, \Sigma_1^{1 / 2} X \Sigma_2^{1 / 2} \right\rangle \tmmathbf{e}_k
  \right\|_2\nn\\
  & \leq & h \left( G'^1, \cdots, G'^m \right) + \sqrt{\sum_k \left\| G^k -
  G'^k \right\|_F^2}
\end{eqnarray}
shows that the Liptchitz constant of $h \left( G^1, \cdots, G^m \right)$ is 1.
Here we have used the triangle inequality for the first inequality, \
Cauchy-Schwarz inequality and $\Sigma_1^{1 / 2} X \Sigma_2 = 1$ on $V \left( r
\right)$ for the last inequality. By concentration of measure in Gauss space \cite{Massart2003concentration},
we obtain
\begin{eqnarray}
  &&\mathbbm{P} \left\{ \left| M \left( r, \mathcal{A}_{\Sigma_1, \Sigma_2}
  \right) -\mathbbm{E}M \left( r, \mathcal{A}_{\Sigma_1, \Sigma_2} \right)
  \right| \geq t \left( r \right) / 2 \right\}\nn\\
   & \leq & 2 \exp \left( - m t^2
  \left( r \right) / 8 \right)
\end{eqnarray}
where $t \left( r \right) := \frac{1}{4} + \frac{\sqrt{\tmop{trace}
\left( \Sigma_1 \right)} + \sqrt{\tmop{trace} \left( \Sigma_2
\right)}}{\sqrt{m}} r$, implying
\begin{eqnarray}
  &&\mathbbm{P} \left\{ M \left( r, \mathcal{A}_{\Sigma_1, \Sigma_2} \right)
  \geq \frac{3 t \left( r \right)}{2} \right\} \nn\\
  & \leq & 2 \exp \left( - m t^2
  \left( r \right) / 8 \right) .
\end{eqnarray}
To complete the proof, we need the following lemma whose proof is based on a
peeling argument:

\begin{lm}[Lemma 3 of \cite{Raskutti2010correlated}]
  \label{lm:peeling}Consider a random objective function $f \left(\tmmathbf{v}; \tmmathbf{u} \right)$ with ${\boldsymbol{u}}$ the underlying
  random variable and $\tmmathbf{v} \in \mathbbm{R}^p$ the variable to be
  optimized with. Suppose that $g : \mathbbm{R}^p \rightarrow \mathbbm{R}_+$
  specifies an increasing constraint function, namely, $\left\{ \tmmathbf{v}:
  g \left( \tmmathbf{v} \right) \leq r \right\} \subseteq \left\{
  \tmmathbf{v}: g \left( \tmmathbf{v} \right) \leq r' \right\}$ for $r \leq
  r'$, $\Omega$ is a non-empty constraint set, and
  \begin{eqnarray*}
    E = \left\{ \tmmathbf{u}: \exists \tmmathbf{v} \in \Omega \tmop{such}
    \tmop{that} f \left( \tmmathbf{v}; \tmmathbf{u} \right) \geq 2 q \left( g
    \left( \tmmathbf{v} \right) \right) \right\},
  \end{eqnarray*}
  where $q : \mathbbm{R} \rightarrow \mathbbm{R}$ is non-negative and strictly
  increasing with $q \left( r \right) \geq \mu$ for all $r \geq 0$. Further
  assume that there exists constant $d > 0$ (which may depend on some other
  parameters) such that
  \begin{eqnarray}
    &&\mathbbm{P} \left\{ \tmmathbf{u}: \sup_{\tmmathbf{v} \in \Omega, g \left(
    \tmmathbf{v} \right) \leq r} f \left( \tmmathbf{v}; \tmmathbf{u} \right)
    \geq q \left( r \right) \right\}\nn\\
     & \leq & 2 \exp \left( - d q^2 \left( r
    \right) \right) .
  \end{eqnarray}
  Then we have
  \begin{eqnarray}
    \mathbbm{P} \left\{ E \right\} & \leq & \frac{2 \exp \left( - 4 d \mu^2
    \right)}{1 - \exp \left( - 4 d \mu^2 \right)} .
  \end{eqnarray}
\end{lm}

We apply Lemma \ref{lm:peeling} to the objective function $f \left( X ;
\mathcal{A}_{\Sigma_1, \Sigma_2} \right) = 1 - \left\| \mathcal{A}_{\Sigma_1,
\Sigma_2} \left( X \right) \right\|_2$, the constraint function $g \left( X
\right) = \left\| X \right\|_{\ast}$, the constraint set $\Omega = \left\{ X :
\left\| \Sigma_1^{1 / 2} X \Sigma_2^{1 / 2} \right\|_F = 1 \right\} $,
$q \left( r \right) = \frac{3}{2} t \left( r \right) = \frac{3}{8} +
\frac{3}{2} \frac{\sqrt{\tmop{trace} \left( \Sigma_1 \right)} +
\sqrt{\tmop{trace} \left( \Sigma_2 \right)}}{\sqrt{m}} r$, and $\mu = 3 / 8$.
Since
\begin{eqnarray}
  &&\mathbbm{P} \left\{ \sup_{g \left( X \right) \leq r, X \in \Omega} f \left(
  X ; \mathcal{A}_{\Sigma_1, \Sigma_2} \right) \geq q \left( r \right)
  \right\}\nn\\
   & \leq & 2 \exp \left( - c m q^2 \left( r \right) \right),
\end{eqnarray}
Lemma \ref{lm:peeling} implies that for properly chosen constant $c > 0$ the event
\begin{eqnarray}
&&\hspace{-1.5cm}E= \{ \exists X \in \Omega \text{\ such that\ } 1 - \left\|\mathcal{A}_{\Sigma_1, \Sigma_2} (X) \right\|_2\nn\\
&&\hspace{-1.5cm} \geq
  \frac{3}{4} + 3 \frac{\sqrt{\tmop{trace} \left( \Sigma_1 \right)} +
  \sqrt{\tmop{trace} \left( \Sigma_2 \right)}}{\sqrt{m}} \left\| X
  \right\|_{\ast}\}
\end{eqnarray}
has probability
\begin{eqnarray}
  \mathbbm{P} \left\{ E\right\} & \leq & 2 \exp \left( - c m \right) .
\end{eqnarray}
So with probability at least $1 - 2 \exp \left( - c m \right)$ we have
\begin{eqnarray}
  &&\hspace{-1cm}\left\| \mathcal{A}_{\Sigma_1, \Sigma_2} \left( X \right) \right\|_2 \nn\\
  &&\hspace{-1cm} \geq
  \frac{1}{4} \left\| \Sigma_1^{1 / 2} X \Sigma_2^{1 / 2} \right\|_F\nn\\
  &&\hspace{-1cm} - 3
  \frac{\sqrt{\tmop{trace} \left( \Sigma_1 \right)} + \sqrt{\tmop{trace}
  \left( \Sigma_2 \right)}}{\sqrt{m}} \left\| X \right\|_{\ast} .
\end{eqnarray}

\section*{Acknowledgment}
The authors would like to thank the reviewers for their helpful suggestions in improving the paper.
\bibliographystyle{IEEEbib}
\bibliography{IEEEabrv,Gongbib}

\begin{IEEEbiographynophoto}{Gongguo Tang}(S'09--M'11) received the B.Sc. degree in mathematics from the Shandong University, China, in 2003, and the M.Sc. degree in systems science from Chinese Academy of Sciences, in 2006, and the Ph.D. degree in electrical and systems engineering from Washington University in St. Louis, St. Louis, MO, in 2011. He is currently a Postdoctoral Research Associate at the Department of Electrical and Computer Engineering, University of Wisconsin-Madison. His research interests are in the area of signal processing, convex optimization, statistics, and their applications.\end{IEEEbiographynophoto}

\begin{IEEEbiographynophoto}{Arye Nehorai}(S'80--M'83--SM'90--F'94) is the Eugene and Martha Lohman Professor and Chair of the Preston M. Green Department of Electrical and Systems Engineering (ESE) at Washington University in St. Louis (WUSTL). He serves as the Director of the Center for Sensor Signal and Information Processing at WUSTL. Earlier, he was a faculty member at Yale University and the University of Illinois at Chicago. He received the B.Sc. and M.Sc. degrees from the Technion, Israel and the Ph.D. from Stanford University, California. Dr. Nehorai served as Editor-in-Chief of IEEE Transactions on Signal Processing from 2000 to 2002. From 2003 to 2005 he was the Vice President of the IEEE Signal Processing Society (SPS), the Chair of the Publications Board, and a member of the Executive Committee of this Society. He was the founding editor of the special columns on Leadership Reflections in IEEE Signal Processing Magazine from 2003 to 2006. He has been a Fellow of the IEEE since 1994 and of the Royal Statistical Society since 1996.\end{IEEEbiographynophoto}
\end{document}